# Contact Engineering High Performance *n*-Type MoTe₂ Transistors

Michal J. Mleczko[1], Andrew C. Yu[1], Christopher M. Smyth[2], Victoria Chen[1], Yong Cheol Shin[1], Sukti Chatterjee[3],  Yi-Chia Tsai[1,4], Yoshio Nishi[1], Robert M. Wallace[2], and Eric Pop[1,5,*]

*1. Department of Electrical Engineering, Stanford University, Stanford CA 94305, USA*

*2. Department of Materials Science & Engineering, Univ. Texas Dallas, Richardson TX 75083, USA*

*3. Applied Materials Inc., Santa Clara CA 95054, USA*

*4. Department of Electrical and Computer Engineering, National Chiao Tung University, Hsinchu 300, Taiwan*

*5. Department of Materials Science & Engineering, Stanford CA 94305, USA*

## Abstract

Semiconducting MoTe₂ is one of the few two-dimensional (2D) materials with a moderate band gap, similar to silicon. However, this material remains underexplored for 2D electronics due to ambient instability and predominantly *p*-type Fermi level pinning at contacts. Here, we demonstrate unipolar *n*-type MoTe₂ transistors with the highest performance to date, including high saturation current (>400 μA/μm at 80 K and >200 μA/μm at 300 K) and relatively low contact resistance (1.2 to 2 kΩ·μm from 80 to 300 K), achieved with Ag contacts and AlOₓ encapsulation. We also investigate other contact metals, extracting their Schottky barrier heights using an analytic subthreshold model. High-resolution X-ray photoelectron spectroscopy reveals that interfacial metal-Te compounds dominate the contact resistance. Among the metals studied, Sc has the lowest work function but is the most reactive, which we counter by inserting monolayer h-BN between MoTe₂ and Sc. These metal-insulator-semiconductor (MIS) contacts partly de-pin the metal Fermi level and lead to the smallest Schottky barrier for electron injection. Overall, this work improves our understanding of *n*-type contacts to 2D materials, an important advance for low-power electronics.

Atomically-thin field-effect transistors (FETs) based on the sulfides and selenides of Mo and W have demonstrated significant current modulation, moderate carrier mobilities,[1, 2] and the capability for low-power complementary logic.[3, 4] In contrast, semiconducting 2H- (α-)MoTe$_2$ remains relatively underexplored, despite moderate indirect band gaps $E_G \approx 0.88$ to 1.0 eV in bulk crystal[5-7] and direct $E_G \approx 1.1$ eV (optical) and 1.2 eV (electronic) in monolayers,[8-10] similar to bulk silicon. Such moderate band gaps may facilitate low-power transistors with tunable injection of either electrons or holes, enabling low-voltage complementary logic and optoelectronics in the visible-to-infrared range.[11, 12] Moreover, 2H-MoTe$_2$ is metastable with a semimetallic 1T' phase, and switching between the two is enabled by temperature,[13] strain,[14] or electrostatic gating,[15] with applications for ultra-low-power switches, phase-change memory, or phase-engineered transistor contacts.[16] Stable metallic nanowire formation has also been recently reported.[17]

Despite these favorable properties, two major challenges have hindered the broader exploration and integration of MoTe$_2$ devices. The first is ambient degradation,[18] as these tellurides are more prone to oxidation than sulfides or selenides,[19] as noted in the rapid oxidation of WTe$_2$ exposed to atmosphere.[20, 21] Chen *et al.*[22] tracked the decomposition of MoTe$_2$ over hours to days in air, which was accompanied by a decline in photoluminescence yield attributed to oxidation around defect sites. The second challenge is that conventional metal contacts to MoTe$_2$ exhibit highly variable Fermi level pinning across reported studies, with poor carrier selectivity and significant Schottky barriers. Previous multi-layer MoTe$_2$ transistors have been predominantly *p*-type[23-26] or ambipolar,[27-33] ostensibly due to contact pinning or minute variations in MoTe$_2$ stoichiometry (including doping from iodine flux agents[23] during crystal growth). Devices with midgap pinning may achieve more unipolar contacts with electrostatic gating or absorbate doping,[28, 34-36] potentially enabling complementary logic, similar to ambipolar WSe$_2$.[3, 4] Besides studies based on initially *n*-type MoTe$_2$ (ostensibly from growth-flux dopants, which diffuse out under thermal treatment),[37, 38] the sole report of unipolar *n*-type transport by Cho *et al.*[16] used laser heating to pattern phase-engineered 1T' contacts. However, such techniques are difficult to scale up, in terms of local thermal budget and laser resolution. Further development of selective, low-resistance contacts is thus required for energy-efficient electronics.

In this work, we address the dual challenges of air sensitivity and contact pinning in few-layer MoTe$_2$, fabricating air-stable, AlO$_x$-encapsulated *n*-type transistors with the highest drive currents reported to date for this layered semiconductor. Our previous method of air-free fabrication,[21, 39] in which channel regions avoid any ambient exposure, is applied to MoTe$_2$ transistors with multiple contact metals. We then obtain self-consistent estimates of electron and hole Schottky barriers, accounting for combined thermionic and tunneling mechanisms of carrier injection.[40] From the conventional metals investigated, Ag contacts yield the lowest *n*-type electron barriers and smallest contact resistance $R_C$, obtained with transfer length measurement (TLM) extractions.[41] $R_C$ appears to be independent of metal deposition pressure, which is explained by the formation of Ag-Te compounds at the MoTe$_2$-metal interface, profiled by high-resolution X-ray photoelectron spectroscopy (XPS). We found higher reactivity with ultra-low work function Sc contacts, which form a disordered metal-telluride complex consuming multiple MoTe$_2$ layers. To mitigate this issue, we insert a chemically-grown monolayer of h-BN[42,43] as a diffusion barrier between Sc and MoTe$_2$, preventing Sc-MoTe$_2$ reactions and helping to de-pin the metal Fermi level. With these metal-insulator-semiconductor (MIS) contacts, we achieve the first completely unipolar *n*-type operation in MoTe$_2$ transistors, demonstrating strongly suppressed reverse leakage current and high-field on-state current saturation.

## Contact Pinning

To first characterize the extent of Fermi level pinning at MoTe$_2$ contacts, we extract effective Schottky barriers from electrical measurements for several common contact metals with work function from ~4.3 to ~5.7 eV (see Table 1). We fabricated transistors using MoTe$_2$ exfoliated from synthetic bulk crystals (Supporting Information, **Section 1**). Following our previous method for passivating oxygen-sensitive few-layer WTe$_2$,[21] we performed all processing in the inert atmosphere of nitrogen gloveboxes (O$_2$, H$_2$O < 3 ppm), such that our MoTe$_2$ was only exposed to ambient air for < 5 minutes prior to contact metal deposition. To protect the channel from oxygen and moisture, we performed metal lift-off in a glovebox connected to an atomic layer deposition (ALD) chamber, allowing us to immediately encapsulate our devices *in situ* with 200 Å of AlO$_x$ via benign, low-temperature (150 °C) ALD.[21] **Figure 1a** presents a schematic of a completed device, with further processing details described in **Methods** and **Section 2** of the Supporting Information.

Theoretical band alignments between our metal contacts and MoTe$_2$ are presented in the left half of **Figure 1b**, based on the Schottky-Mott rule which predicts the "Schottky barrier height" for electron injection to the conduction band, $\Phi_n = \Phi_M - \chi$, where $\Phi_M$ is the metal work function and $\chi$ is the semiconductor electron affinity. This simple theory suggests that preferential electron or hole injection into either conduction or valence bands, respectively, is possible among the examined metals, though with larger Schottky barriers expected for electron injection due to the relatively low electron affinity ($\chi \approx 3.85$ eV) of MoTe$_2$ compared to Si and MoS$_2$.[44,45] However, true band alignments will be determined by the charge neutrality level $E_{CNL}$ established by semiconductor defects and penetrating metal gap states at the contact interface.

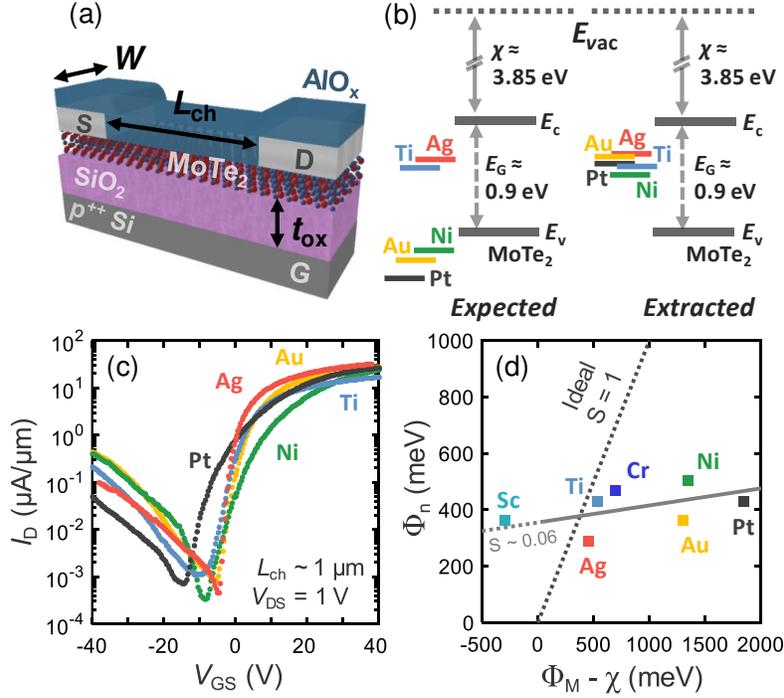

**Figure 1. Contact Pinning** (a) Schematic of AlO$_x$-encapsulated MoTe$_2$ FETs fabricated in inert atmosphere, illustrating channel width ($W$), length ($L_{ch}$), oxide thickness ($t_{ox}$), source (S), drain (D) and back-gate (G) contacts. (b) Expected vs. extracted band alignment of common contact metals to few-layer MoTe$_2$. The extracted band alignments are due to contact Fermi level pinning, as explained in the text. (c) Measured $I_D$ vs. $V_{GS}$ curves of long-channel, ambipolar MoTe$_2$ FETs with five different contact metals (measured at 300 K, on $t_{ox}$ = 90 nm SiO$_2$). (d) Extracted electron Schottky barrier heights to few-layer MoTe$_2$, showing fitted pinning factor $S \sim 0.06$ (solid line), far from the ideal, unpinned case of $S = 1$ (dashed line).

We performed initial electrical measurements of multi-layer devices (channel thickness $t_{ch}$ = 5–10 nm corresponding to 7–14 layers) with 40–60 nm of contact metals (Ag, Au, Ni, Pt, and Ti/Au) deposited in a conventional high-vacuum (HV) evaporator at 1–2 Å/s, with deposition pressures of 0.2 to 5×10$^{-6}$ Torr (with gettering metals like Ti deposited at the lower end of pressure). **Figure 1c** displays measured room-temperature width-normalized drain current density ($I_D$) vs. back-gate voltage ($V_{GS}$) data for long-channel devices ($L_{ch}$ = 0.9–1.1 µm; only reverse sweeps shown for clarity) on 90 nm SiO$_2$ global back-gate oxide at $V_{DS}$ = 1 V. We observe clear ambipolar transport for all contact metals, but with preferential $n$-type conduction by several orders of magnitude, and maximum $I_{on}/I_{off}$ ratios of 10$^4$ to 10$^5$. Qualitatively, these ambipolar transport curves suggest contact pinning at or above mid-gap, consistent with calculations for a charge neutrality level $E_{CNL}$ set deep within the gap by Te-vacancy and possibly Te-interstitial defects at the metal/MoTe$_2$ interface.[46, 47] Our Ti/Au and Ni-contacted samples produce ambipolar transfer characteristics similar to published devices,[27, 30, 32] but with higher $n$-type current densities.

We do not observe any *p*-type dominant transport, even with high work function Pt contacts. This is ostensibly an effect of the AlO$_x$ encapsulation which partly *n*-dopes the MoTe$_2$,[48] comparable to reports of enhanced electron transport in similarly-capped WSe$_2$ and MoTe$_2$ FETs.[3, 32, 36] Similar to these reports,[32, 36] the peak *n*-to-*p* current ratio is enhanced by less than an order of magnitude, although the exact doping amount is difficult to quantify due to significant hysteresis of the MoTe$_2$ devices prior to capping. However, the off-state current minima are not shifted to as far a negative $V_{GS}$ as in prior reports,[36] indicating only moderate doping and enabling study of both current branches. The AlO$_x$ passivation also prevents air exposure, limiting surface oxidation of MoTe$_2$ into MoO$_x$,[49, 50] which could otherwise enhance hole injection (with *p*-type transport correlated to oxygen exposure in uncapped devices).[24, 36, 38]

Prior extractions of effective Schottky barrier heights to MoTe$_2$ used temperature-dependent Arrhenius analysis,[23-25, 29, 30, 33] which assumes reverse leakage current is entirely due to thermionic emission (see Supporting Information, **Section 3**). However, this approach is inaccurate for ambipolar transistors, where the contact Fermi level $E_F$ is pinned deep in the gap, as identified by Das and Appenzeller for WSe$_2$ devices.[51] Deep $E_F$ pinning implies that the reverse current does not reach the exclusively thermionic regime due to significant tunneling injection. Arrhenius analysis will simply interpret this tunneling contribution as thermionic current, underestimating barrier height significantly.[52, 53] This effect may contribute to previous reports of very low barrier heights extracted via conventional Arrhenius analysis for both *p*- and *n*-type injection into MoTe$_2$ with Ti contacts, including $\Phi_p \approx$ 5–130 meV,[23, 25, 27] and $\Phi_n \approx$ 50–190 meV,[30, 33] despite clear evidence of ambipolar transport. For comparison, 50–150 meV Schottky barriers are obtained for unipolar *n*-type MoS$_2$ devices,[41, 52] with contacts well-known to pin just below the conduction band.

Instead, we perform more comprehensive barrier extractions using an analytic Schottky contact model based on Landauer transport theory, developed by Penumatcha *et al.* for ambipolar black phosphorus FETs (more details in **Section 3** of Supporting Information).[40] The subthreshold electron current density is

$$I_{D,n} = \frac{2q}{h} \int_{E_C}^{+\infty} T_C(E) M_C(E) [f_D(E) - f_D(E - qV_{DS})] dE \tag{1}$$

where $T_C(E)$ is the electron tunneling transmission in the Wentzel-Kramers-Brillouin (WKB) approximation, $M_C(E)$ are electron modes in the conduction band, and $f_D(E)$ is the Fermi-Dirac distribution. A similar expression exists for the hole current. This model accounts for combined thermionic and tunneling current, the latter through a simplified WKB model. We extract Schottky barrier heights by using this model to fit electron and hole branches around the current minimum in the subthreshold regime of $I_D$ vs. $V_{GS}$ sweeps for long-channel MoTe$_2$ transistors ($L_{ch} \approx 1$ μm) at low drain bias ($V_{DS} = 100$ meV). This method also self-consistently yields the semiconductor band gap, from the sum of electron and hole Schottky barriers $E_G \approx \Phi_n + \Phi_p$. **Table 1** presents a summary of extracted barrier heights for five contact metals (multilayer samples approaching the electronic bulk value; averaged over multiple extractions and forward/reverse current sweeps). An additional extraction for (Au-capped) Cr, omitted in **Figure 1c** for the sake of clarity, is also included.

Table 1: Extracted Schottky barriers for six contact metals on multi-layer MoTe₂, and reconstructed electronic band gaps (slightly below ~1 eV, as expected[5-7]).

| Contact Metal | Metal Work Function[54] [eV] | $\Phi_n$ [meV] | $\Phi_p$ [meV] | $E_G = \Phi_n + \Phi_p$ [eV] |
|---|---|---|---|---|
| Ag | 4.26 | 290 | 510 | 0.80 |
| Ti | 4.33 | 430 | 480 | 0.91 |
| Cr | 4.5 | 470 | 490 | 0.96 |
| Au | 5.1 | 360 | 580 | 0.94 |
| Ni | 5.15 | 505 | 570 | 1.07 |
| Pt | 5.65 | 430 | 490 | 0.92 |

Ag achieves the smallest room temperature $\Phi_n \approx 290$ meV to multi-layer MoTe₂. However, most $n$-type barriers fall between 400–500 meV, implying deep mid-gap $E_F$ pinning. Qualitatively, these barrier heights are more consistent with the observed ambipolar conduction, unlike previously reported barriers extracted via conventional Arrhenius methodology. Our bulk $E_G$ estimates fall 100–150 meV short of the commonly accepted values,[5-7] likely due to this model's simplifications as well as our underestimates of $\Phi_p$ from suppressed subthreshold hole branches in more unipolar, $n$-type Ag-contacted devices. **Figure 1d** further plots the extracted barrier heights vs. metal work function, along with extractions for pure Sc contacts. We perform a simple linear fit for physically relevant metals (**Table 1** data; $\Phi_M > \chi$) to estimate the pinning factor $S = d\Phi_n/d\Phi_M$. The extracted $S \approx 0.06$ is almost that of a completely pinned material ($S = 0$) rather than the ideal case of the charge neutrality level being set by $\Phi_M$ ($S = 1$, follows Schottky-Mott rule). This is considerably lower than the idealized prediction of $S \approx 0.16$ via density functional theory calculations with metal-induced gap states,[55] though is consistent with a recent report of $S = 0.07$ on ambipolar monolayer MoTe₂ devices (and comparable to that of Ge, $S = 0.05$ but pinned just above the valence band ).[33,56] Clearly additional physical mechanisms beyond the metal work function influence the charge neutrality level, which may include mid-gap chalcogenide defect states (i.e. Te-vacancies or interstitials)[46, 47] or chemical intermixing between reactive MoTe₂ and various contact metals at their interfaces.

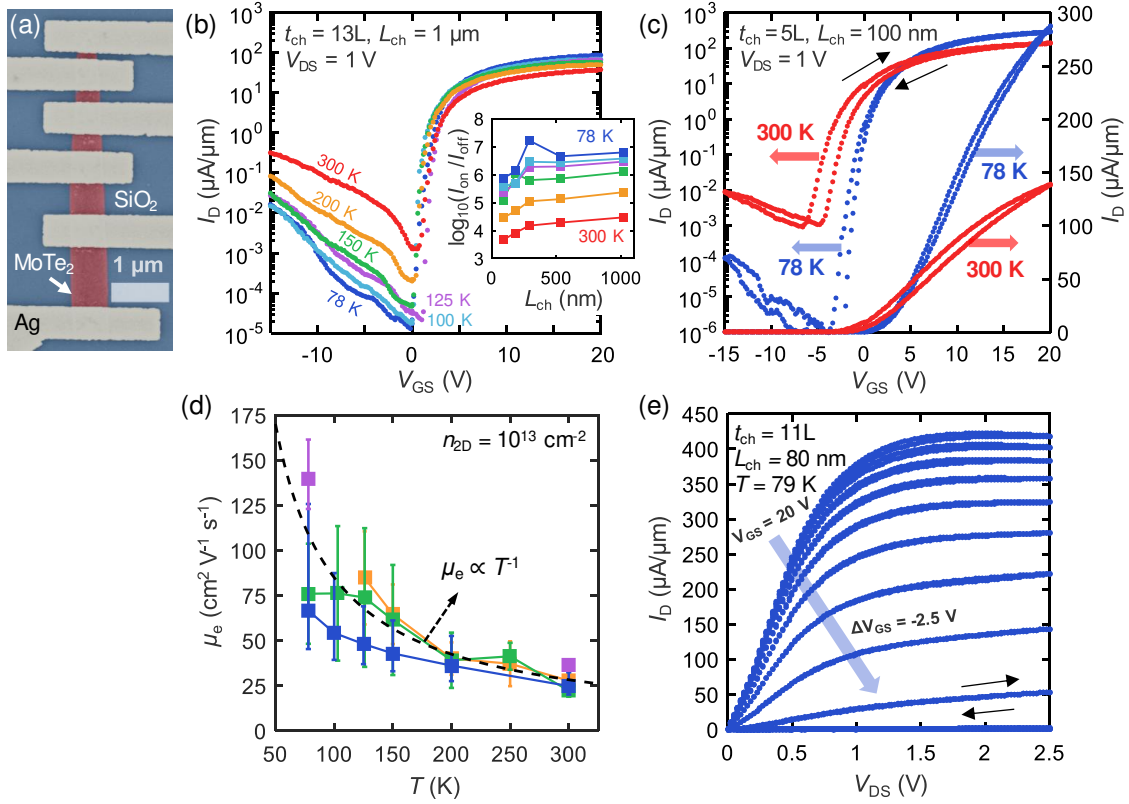

**Figure 2. Ag Contacts** (a) False-colorized SEM micrograph of a typical Ag-contacted MoTe$_2$ TLM test structure. (b) $I_D$ vs. $V_{GS}$ transfer curves of AlO$_x$ encapsulated, Ag-contacted MoTe$_2$ FET on $t_{ox}$ = 30 nm SiO$_2$, measured at 78–300 K. Inset: Peak $I_{on}/I_{off}$ ratio for varying channel lengths fabricated on the same flake across this temperature range. (c) Measured $I_D$ vs. $V_{GS}$ curves of a thin-body, short-channel ($t_{ch}$ = 3.5 nm, $L_{ch}$ = 100 nm) MoTe$_2$ transistor with Ag-contacts ($t_{ox}$ = 30 nm SiO$_2$, dual-sweep from the origin). Black arrows indicate sweep direction and a small amount of hysteresis. (d) TLM-extracted channel electron mobility vs. temperature for several few-layer MoTe$_2$ samples at high carrier density, $n_{2D} \sim 10^{13}$ cm$^{-2}$. Mobility roughly follows a $T^{-1}$ dependence. (e) Measured $I_D$ vs. $V_{DS}$ curves of a short-channel, few-layer Ag-contacted MoTe$_2$ $n$-type FET ($t_{ox}$ = 30 nm SiO$_2$, forward and reverse sweeps, showing almost no hysteresis) demonstrating current saturation density exceeding 420 µA/µm at 79 K ambient temperature. The back-gate voltage decreases from $V_{GS}$ = 20 V (top curve) in -2.5 V steps.

## Silver Contacts

We further investigated Ag electrodes as $n$-type contacts, because they demonstrated the smallest barrier for electron injection as well as the highest current density and steepest subthreshold activation, as shown in **Figure 1c**. We fabricated transistors on 30 nm SiO$_2$ global back-gate oxide (depicted in **Figure 2a**), facilitating induced sheet carrier densities $n_{2D} > 10^{13}$ cm$^{-2}$. In pursuit of clean contact interfaces, we deposited Ag (25 nm, capped with 15 nm Au) in a load-locked, cryopump-driven chamber with evaporation pressures down to 2.5×10$^{-8}$ Torr, just above the crossover to the ultrahigh vacuum regime (UHV, ~10$^{-9}$ Torr and below). **Figure 2b** displays representative $I_D$ vs. $V_{GS}$ transfer curves of a multi-layer MoTe$_2$ FET with Ag contacts (reverse sweep only; negligible hysteresis at positive gate bias) showing predominantly $n$-type transport with current minima around $V_{GS} = 0$. Measurements down to 78 K via closed-loop nitrogen cooling reveal strong temperature-dependence, which is characteristic of a Schottky-barrier-dominated device, including increased on-state current densities at lower temperature due to reduced $R_C$ and enhanced mobility (see discussion below). From suppression of thermionic hole injection at 78 K, we saw three orders of magnitude increase in peak current $I_{on}/I_{off}$ ratio, reaching $10^6$–$10^7$ (**Figure 2b** inset), despite an onset of short-channel effects for $L_{ch} < 100$ nm. We note the stronger temperature-dependence of hole leakage current (at negative $V_{GS}$) above 150 K, indicating more thermionic rather than tunneling charge injection despite the relatively large Schottky barrier (Table 1). This is consistent with expectations of dominant thermionic contribution at higher temperature and low lateral field in the long-channel device of **Figure 2b**, as described in the extraction model and resembling transfer curves of ambipolar black phosphorus transistors.[40] In this regime, the temperature evolution of off-state current resembles that of the strongly thermionic MIS devices presented in the following section.

Per capacitive scale-length theory, ultra-thin body devices should be electrostatically "well-behaved" down to short channel lengths (relative to gate oxide thickness). **Figure 2c** presents both linear and logarithmic transfer characteristics of a thin 5-layer FET ($t_{ch} \approx 3.5$ nm) with a channel length of ~100 nm that maintains predominantly $n$-type transport with peak $I_{on}/I_{off} \approx 10^5$ (~$10^8$) at 300 K (78 K). Forward/reverse sweeps demonstrate low hysteresis in these air-stable $AlO_x$ encapsulated devices. Drive currents double at 78 K to 300 µA/µm at $V_{DS} = 1$ V, surpassing previously highest reported ~100 µA/µm for $p$-type transport in substantially thicker, uncapped $MoTe_2$ samples.[24]

We extract the $MoTe_2$ channel sheet resistance using TLM methodology, from which we estimate intrinsic channel electron mobilities $\mu_e$ (**Figure 2d**).[41] Electron mobilities saturate around 25–36 cm$^2$ V$^{-1}$ s$^{-1}$ at high carrier density at 300 K, matching the range of peak hole mobility from 4-point measurements on intrinsically $p$-type samples.[23] Our electron mobilities rise to 129–137 cm$^2$ V$^{-1}$ s$^{-1}$ at 78–80 K, decaying with a $T^{-1}$ dependence softer than the canonical ~$T^{-1.6}$ evolution expected from homopolar optical phonon scattering under low impurity concentrations.[57] This reduction in temperature coefficient is likely due to encapsulation with our high-$\kappa$ $AlO_x$, where the enhanced dielectric environment dampens low-lying optical phonons, limiting the energetic cross-section for carrier scattering. Short-channel devices experience high-field current saturation for $V_{DS}$ ~ 2.5 V, with record room-temperature current densities >200 µA/µm, increasing by a further ~50% to >400 µA/µm at cryogenic temperatures from reduced $R_C$ and enhanced $\mu_e$ (**Figure S7, S11** in Supporting Information). **Figure 2e** presents a low temperature $I_D$ vs. $V_{DS}$ sweep of a typical few-layer, short channel ($L_{ch}$ ~ 80 nm) device, achieving record saturation current density of 420 µA/µm at 78 K, approaching that of chemically doped or ultra-short channel $MoS_2$.[58] Low-temperature current densities saturate just below 450 µA/µm in bulk samples (Supporting Information, **Section 7**). We note that maximum achievable saturation current densities are most likely limited by device self-heating, and could be further increased in transistors that are better heat sunk, or functioning in pulsed (digital) operation with low duty cycles.[21, 59]

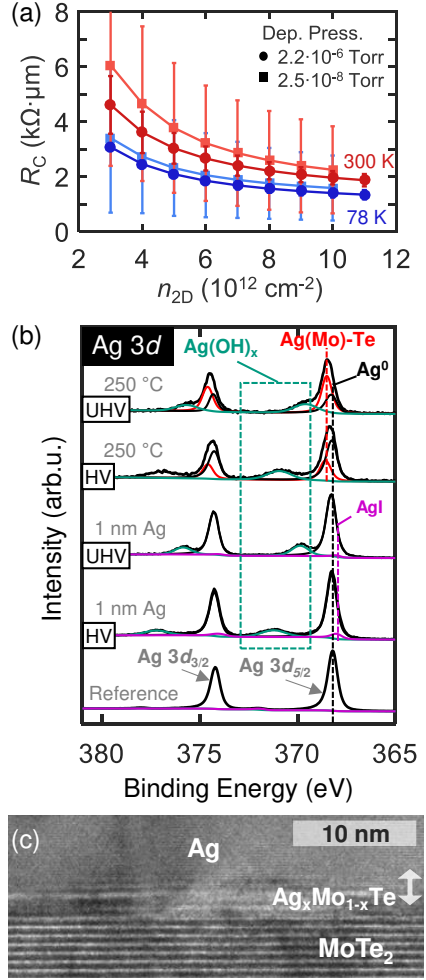

**Figure 3. Ag Contact Chemistry** (a) TLM-extracted contact resistance $R_C$ vs. carrier density at 78 and 300 K for electron injection at Ag-MoTe$_2$ contacts. The extracted $R_C$ does not appear to depend on the metal deposition pressure, unlike in previous work with (less reactive) MoS$_2$ contacts.[41] (b) High-resolution XPS spectra of Ag $3d$ core levels for 1 nm Ag films deposited *in situ* under HV and UHV conditions. Interfacial Ag(Mo)-Te compounds emerge following a 250 °C UHV anneal, which replicates the thermal budget of fabricated devices. (c) High resolution TEM cross-section of an Ag/MoTe$_2$ contact interface, revealing a gradual transition between the metal and layered semiconductor, corresponding to the identified chemical intermixing.

We extract contact resistance via TLM measurements on samples containing 4–6 separate channel lengths, from $L_{ch}$ = 80 nm to over 2 µm. This followed prior methodology demonstrated for MoS$_2$ transistors,[41] extracting $R_C$ from the intercept of net resistance vs. channel length at fixed channel sheet charge density $n_{2D}$ (derived from gate overdrive, accounting for channel quantum capacitance; details in Support-ing Information, **Section 4**). **Figure 3a** presents extracted $R_C$ for two $AlO_x$-encapsulated, multi-layer samples with metal contacts deposited in HV ($2.2 \times 10^{-6}$ Torr, circular symbols) and just above UHV ($2.2 \times 10^{-8}$ Torr, square symbols). Contact resistance ranges from 1.4–1.5 kΩ·µm at 78 K ($n_{2D} \approx 10^{13}$ cm$^{-2}$), and up to ~2.0–2.25 kΩ·µm at 300 K. This is several times higher than the lowest reported $R_C$ for top metal contacts (UHV-deposited Au) on 2H-MoS$_2$ (<750 Ω·µm), though is consistent with projected values from Tsu-Esaki/Arrhenius models assuming comparable Schottky barrier heights on such devices (i.e. for $\Phi_n$ ~ 300 meV for MoTe$_2$ versus the extracted ≤ 150 meV for MoS$_2$).[41] We also note that MoS$_2$ is projected to have a local minimum in $R_C$ around 200 K, attributed to competing effects of reduced thermionic emission (less injected charge from metal into semiconductor) and lower access resistance due to less phonon scattering at lower temperatures.[41] In contrast, our contact resistance to MoTe$_2$ monotonically decreases with cooling down to 78 K, implying that access resistance dominates our contacts, which is consistent with the larger $\Phi_n$ to MoTe$_2$ suppressing the contribution of thermionic emission relative to field emission in the on-state. We extract room-temperature transfer lengths $L_T \approx 120$ nm from TLM analysis (Supporting Information, **Section 4**), a three-fold increase over Au/MoS$_2$ contacts.[41] Our metal contact lengths $L_C = 700$–800 nm $\gg L_T \approx 120$ nm at 300 K and $L_T \approx 380$ nm at 77 K, so our TLM structures and $R_C$ extractions should not be significantly impacted by current crowding.[41]

More unique is the apparent independence of $R_C$ on metal deposition pressure, with $R_C$ curves in **Figure 3a** overlapping within extraction uncertainty, despite two orders of magnitude difference in reactor pressure. This trend nominally persists for UHV-deposited Ag (≤ $5 \times 10^{-9}$ Torr; Supporting Information, **Section 5**). Such invariance contradicts expectations of lower $R_C$ with cleaner interfaces (from lower reactor pressure) for idealized, inert top metal contacts on a van der Waals crystal, suggesting significant chemical modification of the top MoTe$_2$ layer(s) following metallization. However, this is not unexpected given earlier knowledge about MoS$_2$-metal interactions,[60, 61] with the extent and composition of interfacial compounds mediated by both metal-chalcogen reactivity and metal evaporator chamber pressure. In particular, the chemical instability of MoTe$_2$ and persistence of numerous silver-telluride compounds support this conjecture, with prior surface studies of high-temperature Ag nucleation on 1T'- (β-) MoTe$_2$ under UHV detecting covalent bonding and growth of epitaxial Ag$_2$Te islands.[62] However, the 1T' phase of MoTe$_2$ exhibits its greater instability than its 2H allotrope and therefore chemical perturbation by Ag deposition may not manifest analogously on 2H-MoTe$_2$.

To verify such interfacial reactivity, we performed high resolution X-ray photoelectron spectroscopy (XPS) *in situ* on Ag evaporated onto MoTe$_2$ under various reactor pressures, with measured spectra depicted in **Figure 3b** (for experimental details, see Supporting Information, **Section 6**). Despite the inherent Te-rich nature of our MoTe$_2$ (average Te:Mo ratio ≈ 2.3 for MoTe$_2$ discussed here), any reaction products formed between MoTe$_2$ and Ag as deposited at room temperature are below the limit of XPS detection, regardless of reactor base pressure. However, Ag reacts with iodine (residual from crystal growth) to form AgI as evidenced by the chemical states at low binding energy in the corresponding Ag $3d_{5/2}$ (368.10 eV) and I $3d_{5/2}$ (618.8 eV, **Figure S8c,d**) core levels.[63] Iodine is presumably drawn to the Ag-MoTe$_2$ interface during Ag deposition, which is evidenced by the increased iodine concentration detected by XPS after Ag deposition on MoTe$_2$ at room temperature.

Annealing Ag-MoTe$_2$ at 150 °C (temperature of our AlO$_x$ ALD process) drives substantial reactions at the interface with Ag, resulting in the formation of intermetallic Mo$_x$Ag$_{1-x}$Te across multiple MoTe$_2$ layers. Thermally activated intermetallic formation is accompanied by the appearance of an associated chemical state (red curve, 368.55 eV) in the "150°C anneal" Ag $3d$ core level spectrum (**Figure S9**), which persists in the "250°C anneal" Ag $3d$ spectrum (temperature of our post-ALD vacuum anneal, **Figure 3b**). The binding energies of the Mo$_x$Te$_{1-x}$Ag chemical states in the Te $4d$, Mo $3d$, and Ag $3d$ core level spectra relative to bulk MoTe$_2$ and metallic Ag suggest Ag reduces MoTe$_2$, forming a compound with (Mo+Ag):Te ratio of ~1.

Instability of the Ag-MoTe$_2$ interface does not linearly depend on the post-metallization annealing temperature (up to 250 °C). The intensity ratio of metallic Ag and Mo$_x$Ag$_{1-x}$Te chemical states in the corresponding Ag $3d$ core level spectra remain virtually constant (~0.5) in this work. In addition, annealing at 250 °C drives complete dissociation of Ag-I bonds based on the decrease in intensity of AgI chemical states below the limit of XPS detection in the corresponding Ag $3d$ and I $3d_{5/2}$ core level spectra (**Figure S8, S9**). However, it is unclear how the AlO$_x$ encapsulation on the devices fabricated in this work affects iodine diffusion. Nonetheless, the intermetallic formed at the Ag-MoTe$_2$ interface via thermal annealing dominates interfacial chemistry, and thus contact resistance, which may explain why we see no dependence of $R_C$ on metallization pressure. This is in contrast to the case of Au on MoS$_2$, whose interface remains inert even during UHV deposition,[61] exhibiting several times lower $R_C$ for Au contacts deposited at UHV compared to HV due to cleaner interfaces.[41] **Figure 3c** displays high resolution transmission electron micrographs (TEM) of Ag deposited on multilayer MoTe$_2$, confirming the presence of a gradual transition region at this interface, consistent with local intermixing. Partially-visible outlines of MoTe$_2$ layers blend into the Ag film, with spatial inhomogeneity suggesting varying degrees of interfacial reaction mediated by metal grain size and possible MoTe$_2$ defects.

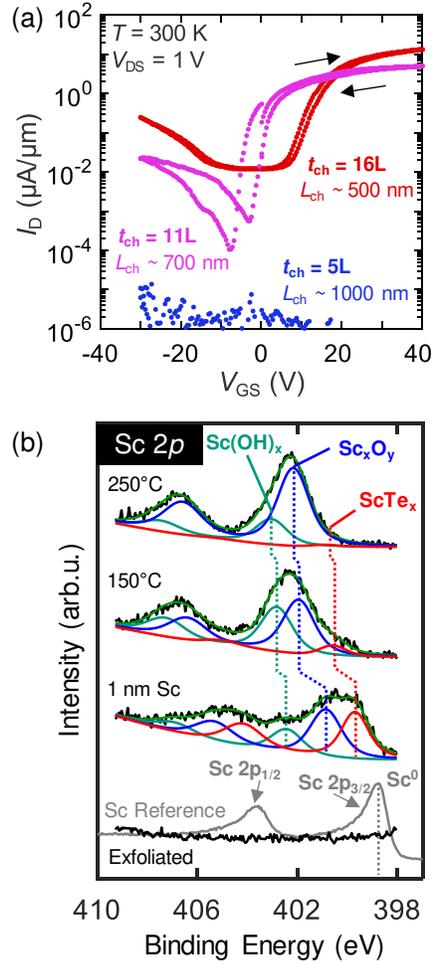

**Figure 4. Sc Contacts** (a) Measured $I_D$ vs. $V_{GS}$ curves of several MoTe$_2$ FETs with Sc metal contacts (on $t_{ox}$ = 90 nm SiO$_2$, dual-sweep from the origin) showing poor, ambipolar performance and non-negligible hysteresis. We do not observe current modulation in thin-body devices (i.e. 5-layers or less). (b) High resolution XPS of Sc $2p$ levels for metal following *in situ* evaporation of 1 nm Sc on MoTe$_2$, and anneals replicating conditions during device fabrication. Spectra on the bottom are from exfoliated MoTe$_2$ prior to processing, as well as pure Sc metal. The green signal overlaid on the measured spectra in black is the fitted envelope. We observe significant ScTe$_x$ formation upon deposition, followed by progressive oxidation of Sc from trace O- and OH- after annealing at 150 °C and 250 °C.

### Scandium Contacts with and without h-BN Insulation

We thus have two unique challenges to realizing *n*-type MoTe$_2$ devices: (1) the low electron affinity of MoTe$_2$, $\chi \approx 3.85$ eV, resulting in a conduction band minimum $E_C$ 150–200 meV above that of Si or MoS$_2$ in bulk,[44, 45] and (2) interfacial compounds from reactions between contact metals and MoTe$_2$, which exhibit unknown work functions and may shift $E_{CNL}$. This first challenge suggests significant Schottky barriers for conventional low work function metals, such as Ag and Ti, even in the ideal theoretical case of **Figure 1b**. However, ultra-low work function metals such as Er and Sc ($\Phi_M \approx 3.1$ and 3.5 eV, respectively) may facil-itate direct charge injection into the conduction band if the contacts are de-pinned. This has been demonstrated by *n*-type conduction in typically *p*-type carbon nanotubes and black phosphorus with Sc or Er electrodes,[64, 65] and significant reduction of Schottky barrier height in Sc-contacted *n*-type $MoS_2$.[52]

To address this first challenge, we evaluated scandium (Sc) contacts to $MoTe_2$. The extreme sensitivity of Sc to trace oxygen necessitated deposition in a custom-built UHV chamber (e-beam evaporation pressure in the low ~$10^{-9}$ Torr, idle base pressure ~$5 \times 10^{-11}$ Torr). 25 nm of Sc was capped with 45 nm Cr and 30 nm Ag, all evaporated at 2.0 Å/s. Subsequent ALD $AlO_x$ capping protects both the $MoTe_2$ and contacting metals from ambient oxidation. Despite the apparent air-stability of encapsulated devices, initial results on multi-layer channels (**Figure 4a**, on 90 nm $SiO_2$ back-gate oxide) were markedly inferior to the conventional metal contacts of **Figure 1**, with lower current densities, increased hysteresis, and ambipolarity suggesting mid-gap pinning. Analytical subthreshold analysis suggests a considerable $\Phi_n \approx 360$ meV for such devices (**Figure 1d**), exceeding values obtained for Ag. Moreover, no significant current above the measurement noise floor was recorded for thin-channel transistors, *i.e.* 5-layers and fewer.

Thus, it appears that poor device performance and the unexpectedly large electron Schottky barrier of Sc contacts to $MoTe_2$ result from severe chemical reaction at this interface, consuming multiple $MoTe_2$ layers.[65] Using XPS, we find that metallic Sc reacts with $MoTe_2$ when deposited at room temperature in either UHV or HV, forming $ScTe_x$, $Sc_xO_y$, and $Sc(OH)_x$, (**Figure 4b**) with peaks at 399.69, 400.86, and 402.42 eV, respectively. Any metallic Sc remaining in the deposited film is below the limit of XPS detection as evidenced by the absence of the expected metallic Sc chemical state at ~398.7 eV.[66] Even in UHV conditions at room temperature, Sc spontaneously reacts with adsorbed species on the $MoTe_2$ surface and background gases within the deposition chamber to form scandium oxide. Fixed charges in oxygen-deficient $Sc_xO_y$ and/or $Sc(OH)_x$ incorporated within the deposited Sc contact could contribute to hysteresis. Formation of Sc oxides and/or hydroxides are thermodynamically favorable (Gibbs free energy $\Delta G^{\circ}_{f,ScxOy}$ = -629.94 kJ/mol, $\Delta G^{\circ}_{f,Sc(OH)x}$ = -411.15 kJ/mol),[67] presumably more favorable than the persistence of Sc-Te bonds (see Supporting Information **Figure S10a** for more details). Note the 0.29 eV shift to lower binding energy exhibited by bulk $MoTe_2$ states detected after annealing at 250 °C (**Figure S10a**). This suggests that a combination of interfacial reaction products formed throughout post metallization annealing dominate the Sc-$MoTe_2$ band alignment. The initial Fermi level position of $MoTe_2$ investigated in studying the Sc-$MoTe_2$ interface indicates *n*-type pinning (0.7 eV above the valence band edge, valence band spectrum not shown in **Figure 4b**). However, the Fermi level shifts to ~0.4 eV above the valence band edge (near mid–gap) after the 250 °C anneal according to shifts exhibited by the $MoTe_2$ chemical states. Therefore, the XPS results are consistent with the ambipolar transport observed in $MoTe_2$ devices employing direct Sc contacts.

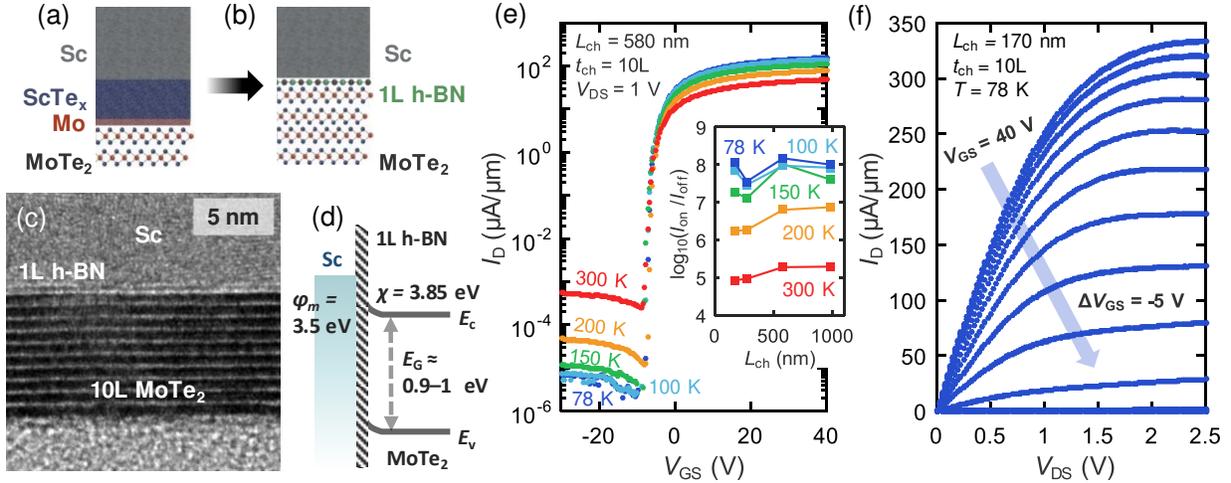

**Figure 5. Sc MIS Contacts with h-BN.** (a) Cartoon of bare Sc on MoTe$_2$, showing interfacial reaction. (b) Cartoon of a 1L h-BN interlayer as a solid-state diffusion barrier between Sc and pristine MoTe$_2$ layers. (c) High resolution TEM micrograph of a h-BN/Sc MIS contact, with underlying MoTe$_2$ thickness consistent with that measured in channel regions. (d) Proposed band alignment for the h-BN/Sc MIS contacts on few-layer MoTe$_2$. (e) Measured $I_D$ vs. $V_{GS}$ curves of a unipolar $n$-type MoTe$_2$ FET with Sc/h-BN MIS contacts (AlO$_x$ encapsulated, back-gated on $t_{ox}$ = 90 nm SiO$_2$). Inset: Peak device $I_{on}/I_{off}$ ratio for various channels fabricated on the same flake. (f) Measured $I_D$ vs. $V_{DS}$ for the shortest channel device on the prior 10-layer sample, demonstrating saturation current densities exceeding 330 µA/µm at 78 K ambient temperature. The back-gate voltage decreases from $V_{GS}$ = 40 V (top curve) in -5 V steps.

In order to limit such extreme interfacial reactions, and more generally de-pin Sc/MoTe$_2$ contacts, we investigated metal-insulator-semiconductor (MIS) contacts.[68-70] The contact metal and semiconductor are physically offset by a nanometer-scale insulating tunnel barrier, limiting the impact of metal-induced gap states and surface dipoles on Fermi level pinning. Although typical insulator layers are often oxides, here we must avoid additional oxygen species and instead we employ hexagonal boron nitride (h-BN) as an atomically-flat, wide-gap insulator ($E_G \approx 6.0$ eV),[71] as shown in **Figure 5b,c**. Recent investigations into the h-BN MIS contact structure have demonstrated improved $R_C$ and reduced Schottky barrier height to MoS$_2$ and black phosphorus FETs,[42, 72, 73] including for nominally-inert Au and Ni electrodes, provided the interlayer thickness is restricted to mono- or bilayers (*i.e.* in the tunneling-dominated limit). For Co electrodes, recent measurements confirm theoretical predictions of an interfacial dipole effect further lowering the effective metal work function by >1 eV in the presence of a h-BN monolayer.[43, 72, 73] Moreover, h-BN may replicate the role of atomically-thin graphene as a metal diffusion barrier[74] and a chemically inert passivation layer for 2D electronics,[71] preventing Sc reactions from scavenging Te in the underlying MoTe$_2$.

Continuous, centimeter-scale h-BN monolayers were grown on re-usable Pt foils via low-pressure CVD (Supporting Information, **Section 8**). These were transferred onto exfoliated MoTe$_2$ flakes on separate 90 nm SiO$_2$ back-gates in a dry, polymer-stamp based process requiring <3 minute air exposure on an 80°C hot plate. After polymer removal and a 200°C anneal for 1–3 hours in nitrogen ambient, the Sc/Cr/Ag contacts and ALD AlO$_x$ capping were deposited by the prior methodology (primarily to prevent oxidation of the Sc-contact stack), followed by a final 250°C vacuum anneal. High resolution TEM (**Figure 5c**) reveals a pristine contact interface, preserving distinct layers with no sign of chemical intermixing. Supporting **Figure S13** confirms consistent MoTe$_2$ thickness between channel and contact regions in thin, few-layer samples. **Figure 5e** displays representative transfer curves of a 10-layer device with Sc/h-BN MIS contacts, demonstrating the most unipolar *n*-type MoTe$_2$ transport to date. The off-state (hole) current is clearly "flattened", indicating strong suppression of hole injection at these contacts.

Steep subthreshold activation indicates significant reduction of the electron Schottky barrier $\Phi_n$, despite electron tunneling through the wide band gap h-BN and the van der Waals gap. Minimum values of inverse subthreshold slope in short-channel devices are almost half the equivalent value in Ag-contacted FETs of similar channel thickness, when adjusted for gate oxide $t_{ox}$ (90 nm vs. 30 nm SiO$_2$). These unipolar MIS-contacted devices maintain $I_{on}/I_{off}$ ratios of $10^5$ at room temperature (~$10^8$ below 80 K) down to electrostatically short channels (e.g. < 200 nm; **Figure 5e** inset), with significantly less ambipolar behavior than Ag-contacted devices in **Figure 2**. **Figure 5f** displays the measured low temperature $I_D$ vs. $V_{DS}$ characteristics, with high-field saturation currents comparable to those with Ag contacts, despite a longer channel and thicker back-gate oxide (90 vs. 30 nm SiO$_2$). The strong suppression ("flattening") of reverse leakage current and excellent high-field saturation represent the most unipolar MoTe$_2$ transistors reported to date. In contrast, previous *n*-type MoTe$_2$ devices formed by electrostatic or molecular doping displayed significant *p*-type reverse leakage as a function of negative $V_{GS}$.[34, 35] Conversely, transistors made from intrinsically *n*-type material operate in depletion mode and barely reach current minima across wide $V_{GS}$ sweeps,[36-38] indicative of a large negative threshold voltage shift rather than de-pinning at the contacts.

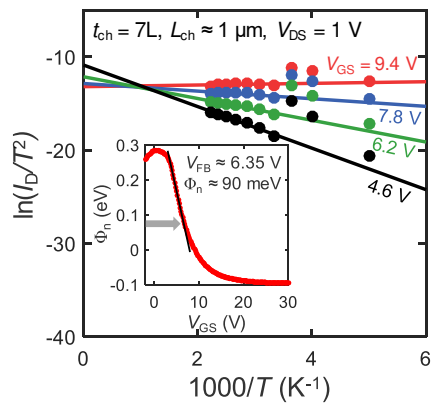

**Figure 6. Sc/h-BN MIS Contacts.** Arrhenius plot of drain current (normalized by width, and by temperature squared) for a long-channel, 7-layer MoTe$_2$ FET with Sc/h-BN contacts on a 90 nm SiO$_2$ back gate oxide, measured at $V_{DS} = 1$ V. Effective Schottky barrier height is extracted assuming the subthreshold drain current is predominantly thermionic emission. Inset: effective electron Schottky barrier extraction for the Sc/h-BN/MoTe$_2$ contact at the flat-band condition, in the high-temperature regime (300 to 450 K). Black line is a linear fit and gray arrow marks the flat-band point where $\Phi_n(V_{GS})$ is no longer linear, yielding $\Phi_n$ ($V_{FB} = 6.35$ V) ≈ 90 meV.

We note that the strongly suppressed off-state (hole) current prevented reliable extraction of Schottky barrier heights by the model described earlier. However, such unipolar behavior suggests significantly de-pinned charge injection, with a predominantly thermionic character enabling conventional Arrhenius extrapolation of $\Phi_n$ (see Supporting Information, **Section 3**). **Figure 6** presents sample fits at various $V_{GS}$ for a $L_{ch}$ ~1 μm, 7-layer unipolar $n$-type $MoTe_2$ device with h-BN/Sc contacts; a good match is achieved for high-temperature data between 300–450 K, for which thermionic and thermally-assisted-tunneling injection is enhanced. An effective, high-temperature electron Schottky barrier height is extracted for the metal/h-BN/van-der-Waals gap system,[42] as determined from $V_{GS} = V_{FB}$, beyond which $\Phi(V_{GS})$ loses linear dependence on gate voltage in conventional barrier models. The inset of **Figure 6** presents such an extraction, with a conservative upper bound of $\Phi_n \approx 90$ meV. We consistently extracted a range of $\Phi_n \approx 80 - 100$ meV across several few-layer, long-channel samples. This represents an effective 200 meV reduction over the average electron barrier of Ag contacts, within the range of Schottky barriers for unipolar $MoS_2$ $n$-FETs ($\Phi_n \leq 150$ meV).[41, 52] Nonetheless, this represents a non-trivial barrier for electron injection, indicating de-pinning is incomplete in light of the theoretical band alignment of Figure 5d. Ultra-low work function metals can move the charge neutrality level of $MoTe_2$ closer to the conduction band, but only when their high reactivity is mitigated by an inert barrier; further study of other metal/h-BN combinations may establish true pinning factors for this semiconductor, wherein the metal work function is not modified by the presence of a local metal-telluride compound.

## Summary

In summary, we demonstrated air-stable, high-performance transistors of semiconducting $MoTe_2$ in the few-layer limit. These were achieved by air-free processing and $AlO_x$ encapsulation, enabling a study of multiple contact metals informed by chemical profiling of metal/$MoTe_2$ interactions at contact interfaces. We achieved highest performance with Ag contacts despite (or perhaps because of) the formation of an Ag-Te contact interlayer, achieving record current densities >400 μA/μm at 80 K. We also achieved the most unipolar $n$-type devices with Sc contacts employing a h-BN contact interlayer. This was required to prevent interfacial Sc-Te reactions, also functioning as a MIS tunnel barrier to partially de-pin the Fermi level. Together, these are the highest performance unipolar $n$-type $MoTe_2$ transistors demonstrated to date, complementary to prior published $p$-type $MoTe_2$ devices. More generally, we found strong evidence of metal-chalcogen reactivity as a key engineering parameter in contact design, demonstrating strategies for both the exploitation and prevention of such reactions at these interfaces.

**Methods**

We exfoliated MoTe$_2$ flakes from synthetic bulk crystals, grown by Chemical Vapor Transport with a molecular precursor source (see Supporting Information, **Section 1**), onto SiO$_2$/p$^{++}$ Si substrates with oxide thickness $t_{ox}$ = 30 (for some Ag-contacted devices) or 90 nm (other contacts). Following our previous method for passivating oxygen-sensitive chalcogenides,[21, 39] we performed all processing in the inert atmosphere of nitrogen gloveboxes (O$_2$, H$_2$O < 3 ppm), coating samples in a protective PMMA layer that also served as resist for electron beam (e-beam) lithography. Following patterning of top-contacts, we developed our samples in air and quickly transferred them to e-beam metal evaporators, limiting the exposure of our contact regions to ambient atmosphere to only 1 to 5 minutes. To prevent channel oxidation, we performed metal lift-off in a nitrogen glovebox followed by *in situ* encapsulation with 200 Å of AlO$_x$ via benign, low-temperature atomic layer deposition (ALD; 150°C thermal process using H$_2$O and trimethylaluminum precursor) to act as an oxygen and moisture barrier. We perform a final 250°C anneal in our vacuum probe station (~10$^{-5}$ torr) to suppress hysteresis in electrical measurements. Devices were measured in a Janis Cryogenic probe station with a Keithley 4200-SCS parameter analyzer. Further processing details are outlined in **Section 2** of the Supporting Information, including Raman spectra of encapsulated flakes. XPS analysis and sample preparation are described in detail in **Section 6** of the Supporting Information. h-BN growth, transfer and characterization are covered in **Section 8**. TEM cross-sections were prepared by Evans Analytical Group using a FEI Dual Beam FIB/SEM and both Tecnai TF-20 and Tecnai Osiris FEG/TEM units at 200 kV.

**ASSOCIATED CONTENT**

**Supporting Information**

The Supporting Information is available free of charge on the ACS Publications website at DOI: XXXXX. Bulk crystal growth and characterization, Raman spectra. Full details of analytical Schottky barrier model and transfer length measurement fitting. Details of XPS sample preparation and measurements with supplementary spectra for Ag/MoTe$_2$ and Sc/MoTe$_2$ interactions. Additional transistor saturation curves. Methods of h-BN film growth, transfer and characterization.

**AUTHOR INFORMATION**

**Corresponding Author:** *E-mail: epop@stanford.edu

**Present Affiliations:** Intel Corporation (M.J.M); Department of Electrical Engineering and Computer Science, Massachusetts Institute of Technology (A.C.Y.); Department of Electrical and Computer Engineering, University of Illinois at Urbana-Champaign (Y.-C.T.).

**Notes:** The authors declare no competing financial interest.

**ACKNOWLEDGEMENTS**

Work was completed at the Stanford Nanofabrication Facility (SNF) and Stanford Shared Nano Facilities (SNSF). The authors would like to acknowledge support from H.J. Silverstein, H.H. Kuo and I.R. Fisher in bulk crystal growth, as well as technical assistance from C.D. English (TLM analysis), B. Magyari-Köpe (DFT analysis), C.M. Neumann, K.K.H. Smithe and I.M. Datye (substrate preparation), and Hui Zhu (STM/XPS analysis). We are also grateful for conversations with H.-S.P Wong and K. Saraswat. Work was sponsored in part by the Air Force Office of Scientific Research (AFOSR) grant FA9550-14-1-0251, the National Science Foundation EFRI 2-DARE grant 1542883, Army Research Office grant W911NF-15-1-0570, the Stanford SystemX Alliance, and the US/Ireland R&D Partnership (UNITE) under the NSF award ECCS-1407765. TEM work was sponsored by the Applied Materials corporation. M.J.M. would like to acknowledge an NSERC PGS-D fellowship. RMW and EP acknowledge the support of the NEWLIMITS center in nCORE, a Semiconductor Research Corporation (SRC) program sponsored by NIST through award number 70NANB17H041, and of ASCENT, one of six centers in JUMP, a SRC program sponsored by DARPA.

## ToC Figure:

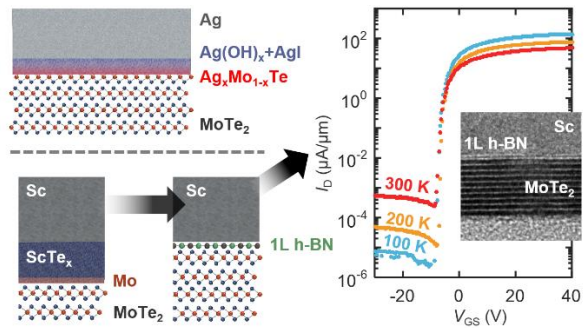

# Supporting Information

## Contact Engineering High Performance *n*-Type MoTe₂ Transistors

Michal J. Mleczko[1], Andrew C. Yu[1], Christopher M. Smyth[2], Victoria Chen[1], Yong Cheol Shin[1], Sukti Chatterjee[3], Yi-Chia Tsai[1,4], Yoshio Nishi[1], Robert M. Wallace[2], and Eric Pop[1,5,*]

*1. Dept. of Electrical Engineering, Stanford University, Stanford CA 94305, USA*

*2. Dept. of Materials Science & Engineering, Univ. Texas Dallas, Richardson TX 75083, USA*

*3. Applied Materials Inc., Santa Clara CA 95054, USA*

*4. Dept. of Electrical and Computer Engineering, National Chiao Tung University, Hsinchu 300, Taiwan*

*5. Dept. of Materials Science & Engineering, Stanford CA 94305, USA*

*[*]Contact: epop@stanford.edu*

## 1. Bulk Crystal Growth

Mirroring a technique previously applied towards facile synthesis of WTe₂ crystals, bulk MoTe₂ exfoliation sources were grown by re-crystallizing a commercial molecular powder using closed-tube chemical-vapor transport (CVT).[1] Molybdenum ditelluride powder (ESPI Metals MoTe₂, 99.9%) was sealed in quartz ampoules with elemental iodine as a transport agent (Alfa Aesar, 99.99+%) at 5 mg/cm³, evacuated under argon. To remain below the ~850–900 °C transition range for the 1T'-/β- semimetallic polytype,[2, 3] the central hot zone was kept at 800 °C, maintaining a 100 °C thermal gradient along a ~11 cm transport length during 14 days of growth. Despite a lower base temperature than WTe₂ CVT, a high yield of millimetric crystal platelets was achieved across multiple growths (**Figure S1a**), with layered structure evident in scanning electron microscopy (SEM) micrographs of sheet-like gradations in edge terraces (**Figure S1b**). Scanning tunneling microscopy (STM) studies confirmed crystal composition and quality consistent with commercially-available synthetic samples, with trace levels of silicon and iodine near bulk surfaces incorporated from the growth ampoule and flux agent, respectively.

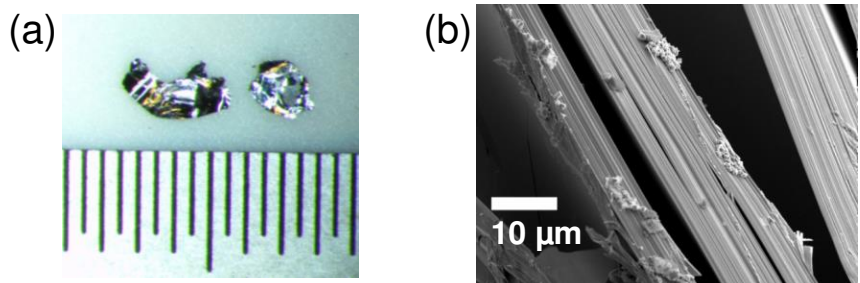

**Figure S1:** (a) CVT-grown bulk MoTe₂ crystals with mm increments for scale.
(b) SEM micrograph displaying layered structure at edges.

## 2. Air-Free Device Fabrication and Characterization

We fabricated top-contacted, back-gated MoTe$_2$ field-effect transistors (FETs) avoiding any air exposure to channel regions, as outlined in a prior study on encapsulated WTe$_2$.[1] We exfoliated flakes onto p$^{++}$ Si chips with 30 or 90 nm SiO$_2$ (dry thermal growth) in an N$_2$ glovebox with O$_2$, H$_2$O < 1 ppm using a low-residue thermal release tape (NittoDenko Revalpha series). After an acetone/2-propanol solvent bath, chips were coated *in situ* with a 300 nm layer of 950k polymethyl methacrylate (PMMA) (MicroChem A5), acting as a protective film while optically searching for suitable flakes to make devices, and as a resist layer for a two-step electron-beam lithography process (patterning alignment markers then contacts in the same resist layer; Raith 150 at 20 kV). We developed contacts in open air and rapidly transferred chips to any of several electron-beam metal evaporators, with deposition pressures spanning the high vacuum (HV) to ultra-high vacuum (UHV) range as outlined in the manuscript; only contact regions saw ambient atmosphere, for sub-5 minute periods.

Following metallization, we performed lift-off by acetone/2-propanol soaking in another N$_2$ glovebox (O$_2$, H$_2$O < 3 ppm) interfacing a Savannah thermal atomic layer deposition (ALD) reactor used to immediately encapsulate devices with 200 Å of amorphous AlO$_x$. To minimize TMD oxidation during ALD, AlO$_x$ growth was conducted at a lower temperature (150 °C), using a less reactive H$_2$O reagent, after first saturating surfaces with 10 leading pulses of trimethylaluminum (TMA) metal precursor. Finally, we vacuum-annealed devices for 1 hour at 250 °C in a Janis Cryogenic probe station, cooling to ambient over several hours under vacuum levels below 5×10$^{-5}$ Torr prior to electrical measurement. Our FETs demonstrate stable performance over weeks of storage.

Smooth nucleation of the capping alumina allowed us to determine MoTe$_2$ channel layer-count from atomic force microscopy (AFM) topography profiles (Veeco Dimension 3100 in soft-tapping mode). We determine layer-count from channel thickness assuming atomic interlayer spacing of ~0.7 nm with an extra ~0.2–0.3 nm offset for the first layer consistent with recent reports and bulk lattice constants.[4,5] Optical transparency of our alumina capping enabled us to verify layer-specific vibrational Raman modes, as presented in **Figure S2** for encapsulated samples, collected using a 1.25 mW, 532 nm laser (Horiba LabRam; renormalized to peak intensity). Laser spot-size was confined to several-µm, with no oxide or flake damage detected by subsequent optical and AFM profiling.

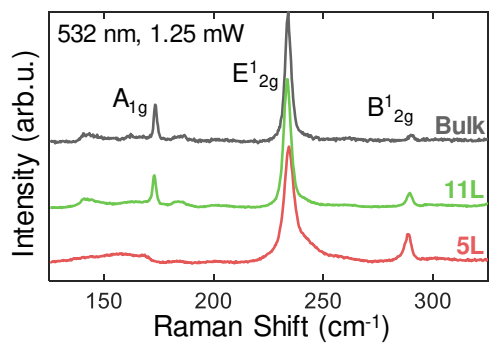

**Figure S2:** Raman spectra of AlO$_x$-capped MoTe$_2$ flakes showing preservation of characteristic vibrational peaks, as well as novel persistence of a cross-plane B$^1_{2g}$ mode in bulk samples (Horiba LabRam, 532 nm laser, 1.25 mW).

Both characteristic cross-plane $A_{1g}$ and in-plane $E^1_{2g}$ modes appear at 173 and 234 cm$^{-1}$ respectively, matching published studies on freshly-exfoliated few-layer crystals,[4, 6] undergoing minimal softening or stiffening (<1 cm$^{-1}$) approaching 5 layers (5L) as in previous reports. This includes reduction of $A_{1g}$ intensity in the 5-layer limit ($t_{ch} \approx 3.5$ nm) corresponding to peak-splitting recorded across 3–10L samples.

An additional mode around 289–291 cm$^{-1}$ matches reports of a cross-plane $B^1_{2g}$ vibration, in which metal and tellurium atoms oscillate in opposite directions perpendicular to the layer plane.[4] Thought to be inactive in both monolayer and bulk, it emerges with increasing relative intensity down to bilayer thickness, matching the general trend in **Figure S2**. A novel persistence of this mode is recorded here in bulk capped samples (tens to hundreds of nanometers thick), albeit with low intensity. An absence of surface disorder from ambient oxidation may contribute to this effect, as suggested by the rapid disappearance of Raman peaks in exposed WTe$_2$,[7] alongside an enhanced local dielectric constant under the AlO$_x$. Similar activation of nominally bulk-inactive modes has been observed in WSe$_2$/h-BN stacks,[8] attributed to electron-phonon coupling at interfaces, suggesting an influence of encapsulating layers on the Raman spectra of 2D crystals.

## 3. Schottky Barrier Extraction

Schottky barrier extraction from 2D FETs has conventionally assumed reverse leakage current (i.e. *p*-type transport for an *n*-FET) is dominated by thermionic emission of a single carrier type.[9-13] Arrhenius analysis of thermionic emission is then used to estimate the effective Schottky barrier. Current density per unit width from thermionic emission at a metal-semiconductor junction is modelled as:[14]

$$I_D = A^* T^2 e^{-\frac{q\Phi_B}{k_B T}} e^{\frac{qV}{n k_B T}} \left( 1 - e^{-\frac{qV}{k_B T}} \right)$$ (S.1)

where $q$ is the unit electron charge, $k_B$ is the Boltzmann constant, $T$ is temperature, $A^*$ is the effective Richardson's constant ($A^* = 4\pi m^* q k_B^2/h^3$), $\Phi_B$ is the effective barrier height from the metal Fermi energy to semiconductor conduction band, $V$ is the voltage applied at the junction, and $n$ is the metal-semiconductor ideality factor (accounting for non-idealities of the Schottky barrier height such as image force lowering).

For a MOSFET in subthreshold, the voltage dropped at the metal-semiconductor contact is roughly the drain bias, $V \approx -V_{DS}$. $\Phi_B$ is furthermore a function of gate-source voltage, $V_{GS}$. Multiplying together the terms with exponential $qV/k_B T$ and assuming an ideality factor $n \approx 1$, we can rewrite (S.1) in the form commonly used to model the subthreshold thermionic emission in a MOSFET:

$$I_D = A^* T^2 e^{-\frac{q\Phi(V_{GS})}{k_B T}} \left( 1 - e^{-\frac{qV_{DS}}{k_B T}} \right).$$ (S.2)

The polarity of $I_D$ is arbitrarily flipped here as (S.1) is current flowing out while (S.2) is now current flowing into the contact. To extract effective Schottky barrier height, the drain voltage is biased such that $qV_{DS} \gg k_B T$ so $[1 - \exp(-qV_{DS}/k_B T)] \approx 1$ across the range of interest for $T$. Then taking the natural log of $I_D/T^2$ yields an equation linear with $1/T$, where the slope is proportional to the effective Schottky barrier height $\Phi(V_{GS})$ at a specific gate bias:

$$\log\left(\frac{I_D}{T^2}\right) = \log\left(A^*\right) - \left[\frac{q\Phi\left(V_{GS}\right)}{k_B}\right]\left(\frac{1}{T}\right) \tag{S.3}$$

The true Schottky barrier height $\Phi_{SB}$ is obtained at the flat-band bias, $V_{GS} = V_{FB}$, beyond which $\Phi(V_{GS})$ is no longer linear with gate voltage due to non-negligible tunneling contributions to the reverse current.

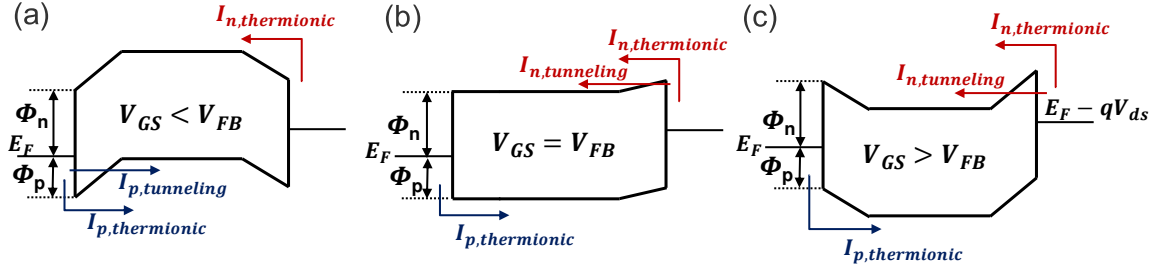

**Figure S3:** Band alignment of an ambipolar Schottky barrier FET (a) below, (b) at, and (c) above flat-band biasing in gate voltage with principal mechanisms of carrier injection and true barrier heights $\Phi_n$, $\Phi_p$ demarcated for a contact Fermi level pinned deep within the band gap.

In the case of an ambipolar transistor, charge-neutrality level pinning sufficiently deep in the band gap implies that reverse current will never reach the exclusively thermionic regime, and it will instead be dominated by tunneling injection. This is outlined in the band diagrams in **Figure S3** across various gate biasing regimes, demonstrating joint contributions of thermionic and tunneling currents to electron/hole barrier height $\Phi_n$ / $\Phi_p$. Arrhenius analysis will simply incorporate the tunneling contribution into a thermionic calculation, underestimating barrier height significantly.[15]

We therefore chose to perform more comprehensive barrier extractions using an analytic Schottky model based on Landauer transport theory, recently developed for ambipolar, 2D black phosphorus FETs.[16] This model assumes no voltage drop across the channel and that the current characteristics are dominated by the contacts. Fits are made using the subthreshold regime of $I_D$ vs. $V_{GS}$ sweeps of MoTe$_2$ transistors, centered on the point of minimum off-current. Width-normalized hole and electron contributions to current are defined by:

$$I_{D,p} = \frac{2q}{h} \int_{-\infty}^{E_V} T_V(E) M_V(E)[f_D(E) - f_D(E - qV_{DS})]dE \tag{S.4}$$

$$I_{D,n} = \frac{2q}{h} \int_{E_C}^{+\infty} T_C(E) M_C(E)[f_D(E) - f_D(E - qV_{DS})]dE \tag{S.5}$$

These integrate energy $E$ relative to conduction and valence band edges, $E_C$ and $E_V$. $T(E)$ denotes carrier tunneling transmission in the Wentzel-Kramers-Brillouin (WKB) approximation, $f_D(E)$ is the Fermi-Dirac distribution, and $M(E)$ are electron/hole modes in conduction/valence bands. Modes are defined by:

$$M_V(E) = \frac{g_v}{\pi\hbar}\sqrt{2m_h^*[E_V(V_{GS}) - E]}, \ E_V(V_{GS}) > E \tag{S.6}$$

$$M_C(E) = \frac{g_v}{\pi\hbar}\sqrt{2m_e^*[E - E_C(V_{GS})]}, \ E > E_C(V_{GS}), \tag{S.7}$$

where $g_v$ is the valley degeneracy, $m_e^*$ and $m_h^*$ are carrier effective masses, and $E_C(V_{GS})$ and $E_V(V_{GS})$ are the gate-bias dependent band positions in the channel. WKB tunneling transmission is given by $T(E) = T_S T_D / (1 - R_S R_D)$, where $T_{S(D)}$ and $R_{S(D)} = 1 - T_{S(D)}$ are the respective transmission and reflection through individual source (drain) Schottky barriers. These are calculated assuming triangular band profiles at the metal-semiconductor contact across a characteristic length $\lambda = [(\varepsilon_{ch}/\varepsilon_{ox})t_{ch}t_{ox}]^{1/2}$ into the channel, with:

$$T_{S(D),WKB} = \exp\left(-\int_0^\lambda \kappa(E)dx\right) \tag{S.8}$$

Where $\kappa_V(E) = \frac{1}{\hbar}\sqrt{2m_h^*[E - E_V(x)]}$ and $\kappa_C(E) = \frac{1}{\hbar}\sqrt{2m_e^*[E_C(x) - E]}$ defining complex carrier momentum along the valence and conduction bands. Both thermionic and tunneling injection of electrons and holes are thus considered across subthreshold sweeps.

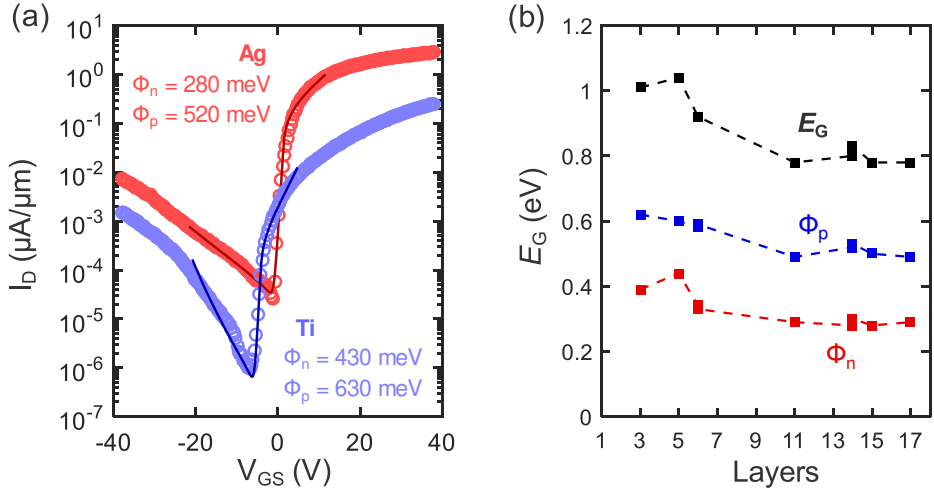

**Figure S4:** (a) Representative transfer curves of long-channel encapsulated MoTe₂ FETs (300 K, $V_{DS} = 100$ mV, on 90 nm SiO₂) depicting subthreshold fits of an analytical Schottky barrier model (lines) against experimental data (symbols) for Ag and Ti contacts. (b) Extracted electron, hole Schottky barrier heights and electronic band gap $E_G$ vs. layer count for Ag-contacted MoTe₂ transistors. Experimentally reconstructed $E_G$ increases with decreasing layer count as expected, but the values are 100-150 meV lower than expected.

We simultaneously fit both $\Phi_n$ and $\Phi_p$ for ~1 μm long, AlO$_x$-capped MoTe₂ FETs on 90 nm SiO₂, at low drain bias $V_{DS} = 100$ mV. In-plane bulk MoTe₂ carrier masses of $m_e^* \approx 0.49m_0$ and $m_h^* \approx 0.61m_0$ were used, extracted from parabolic fits to a DFT band model, calculated in VASP PBE with spin-orbit coupling. Sample fits for Ti and Ag are shown in **Figure S4a**, in good agreement with subthreshold current data.

**Figure S4b** presents the reconstructed band gap values $E_G \approx \Phi_n + \Phi_p$ from Schottky barrier fits to silver-contacted devices of varying thickness. This model generally captures the trend of increasing electronic band gap with reduced layer count, albeit with a 100-150 meV underestimate of established values (~0.9 to 1.2 eV). For more $n$-type Ag-contacted devices, however, gradual subthreshold $p$-type activation may produce underestimates of $\Phi_p$ (with fitting to this weaker branch dominating uncertainty). Local variation in layer number across a sample may also have a smaller contribute to extraction uncertainty, because we selected channel regions of a consistent thickness, as validated by optical imaging and high-resolution AFM and SEM profiling.

**Figure S5** presents a summary of extracted barrier heights and the pinning factor $S = d\Phi_n/d\Phi_M$ from **Figure 1d** in the manuscript, compared to prior published Arrhenius-based estimates.[9-11, 13, 17] This comparison reveals the prevalence of $p$-type transport in uncapped MoTe$_2$ devices. Additionally, it demonstrates consistent underestimates of true barrier height under the purely thermionic contact models employed in other studies, with reported $\Phi_n$ or $\Phi_p$ much closer to band edges despite clearly ambipolar FET transport.

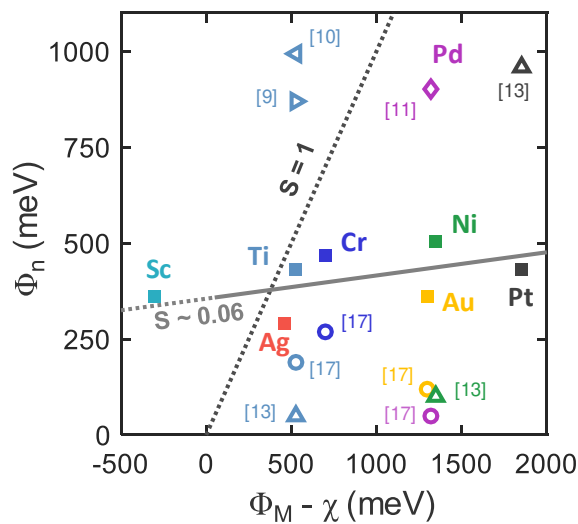

**Figure S5:** Summary of estimated electron Schottky barrier heights for common contact metals on MoTe$_2$, contrasting our values (analytical subthreshold model; solid squares) against prior estimates using Arrhenius analysis (open symbols). A fit of pinning factor ($S$) for our analytically fitted barrier heights is provided.

## 4. Transfer Length Measurement Analysis

We use Transfer Length Measurement (TLM) analysis to simultaneously measure contact resistance and intrinsic mobility in MoTe$_2$ FETs, using multiple patterned channel lengths $L_{ch}$ (minimum 4-6 channels, from 80 nm to >2 µm in length). We use average channel widths for all calculations, determined from post-measurement SEM imaging. Following a methodology established for exfoliated MoS$_2$ FETs,[18] we model width-normalized net channel resistance $R_{TOT}$ as a linear function of channel length $L_{ch}$:

$$R_{TOT}\left(L_{ch}\right) = V_{DS}\big/I_D = L_{ch}R_S + 2R_C \qquad (S.9)$$

Sheet resistance $R_S$ is extracted from the linear slope and contact resistance $R_C$ from half of the $R_{TOT}$-intercept in the TLM plots, such as **Figure S6c**.

We perform extractions at fixed 2D sheet charge density, accounting for quantum capacitance of the channel ($C_{dq}$) in series with the gate oxide capacitance ($C_{ox}$). At high carrier density, the channel 2D electron sheet density is given by:[19]

$$n_{2D} = \frac{1}{q} \frac{C_{ox} C_{dq}}{C_{ox} + C_{dq}} \left( V_{GS} - V_T - V_{crit} \right),$$

(S.10)

$$V_{crit} = \frac{E_G}{2q} + \frac{k_B T}{q} \ln \left( \frac{C_{ox}}{C_{dq} + C_{ox}} \right) + \frac{k_B T}{q} \frac{C_{dq}}{C_{ox}} \ln \left( \frac{C_{ox}}{C_{dq} + C_{ox}} \right)$$

where $V_T$ is device threshold voltage, $V_{crit}$ is the critical gate voltage above which this approximation is valid, $C_{ox} \approx 116$ nF/cm$^2$ for 30 nm SiO$_2$ is the experimentally-verified capacitance (per area) of the global back gate, and $C_{dq}$ is the degenerate limit of the channel quantum capacitance. $C_{dq}$ is given by:

$$C_{dq} = q^2 g_{2D} \approx q^2 \frac{g_S g_V m_e^*}{2\pi \hbar}$$

(S.11)

$g_V = 6$ is the valley degeneracy for few-layer MoTe$_2$, and $m_e^* \approx 0.49 m_0$ is the density of states effective mass. We calculate $C_{dq} \approx 197$ μF/cm$^2$, over ~1000x larger than $C_{ox}$, and $V_{crit} \approx 0.33$ V. The conventional expression for $n_{2D}$ (no series $C_{dq}$ and $V_{crit} = 0$) is 30% larger at $10^{12}$ cm$^{-2}$ but only 2.5% larger at $10^{13}$ cm$^{-2}$. Thus, our quantum capacitance correction is less than for single layer MoS$_2$ because few-layer MoTe$_2$ has a smaller band gap and larger valley degeneracy,[19] which increases $C_{dq}$ and reduces $V_{crit}$. Furthermore, we perform all extractions at high carrier density (close to $n_{2D} \approx 10^{13}$ cm$^{-2}$) where quantum capacitance has a minimal effect.

Conventional bulk MOSFET theory poorly models $V_T$ in our devices due to their ultra-thin floating bodies. Traps and fixed charge in our AlO$_x$ encapsulation further complicate the observed $V_T$. As such, we model drain current with the most simplified first-order model of the linear $V_{GS}$ region[14]

$$I_D = \mu_{eff} C_{ox} \left( V_{GS} - V_T \right) V_{DS}.$$

(S.12)

This form is best for experimental fitting because it allows $V_T$ and effective mobility $\mu_{eff}$ to act as fitting parameters, with fewer assumptions about precise channel physics. $V_T$ is estimated by linearly extrapolating $I_D(V_{GS})$ to the $V_{GS}$-axis intercept, using a line tangent to the point of max transconductance $\partial I_D / \partial V_{GS}$ (using the reverse sweep). **Figure S6c** presents linear fits of width-normalized channel resistance at varying overdrive voltages for a 5-channel TLM sample (11-layer thick MoTe$_2$) at room temperature. Both $R_C$ and $R_S$ are plotted in **Figure S6d** and **S6e**, with error bars delineating 95% confidence intervals of the linear least squares fit. Extracted values saturate in the high carrier density limit ($n_{2D} \geq 10^{13}$ cm$^{-2}$).

2D sheet resistance provides the basis for an intrinsic carrier mobility,[20] independent of contact effects:

$$\mu_i = \frac{1}{qR_S n_{2D}} \qquad (S.13)$$

Converting $R_C$ to a specific contact resistivity $\rho_C$ enables extraction of the transfer length $L_T$, the characteristic length over which injected current decays into a contact:

$$R_C = \frac{\rho_C}{L_T} \coth\left(\frac{L_C}{L_T}\right) \qquad (S.14)$$

$$L_T = \sqrt{\frac{\rho_C}{R_S}}, \qquad (S.15)$$

where $L_C$ is the mean contact length, varied between 500-700 nm across devices. Sample extractions for a representative MoTe$_2$ device in **Figure S6h** reveal room temperature transfer lengths up to $L_T \approx 115$ nm, representing nearly 3x increase over comparable Au/MoS$_2$ contacts ($L_T \approx 40$ nm).[18] This value increases to over 300 nm below 80 K, nearly matching that of Au/MoS$_2$ contacts in this lower temperature regime. However, there was a constant ~40% uncertainty in our $L_T$ extractions. As $L_T < L_C$ across our entire measurement temperature range, we can assume current crowding is negligible in our electrical measurements.

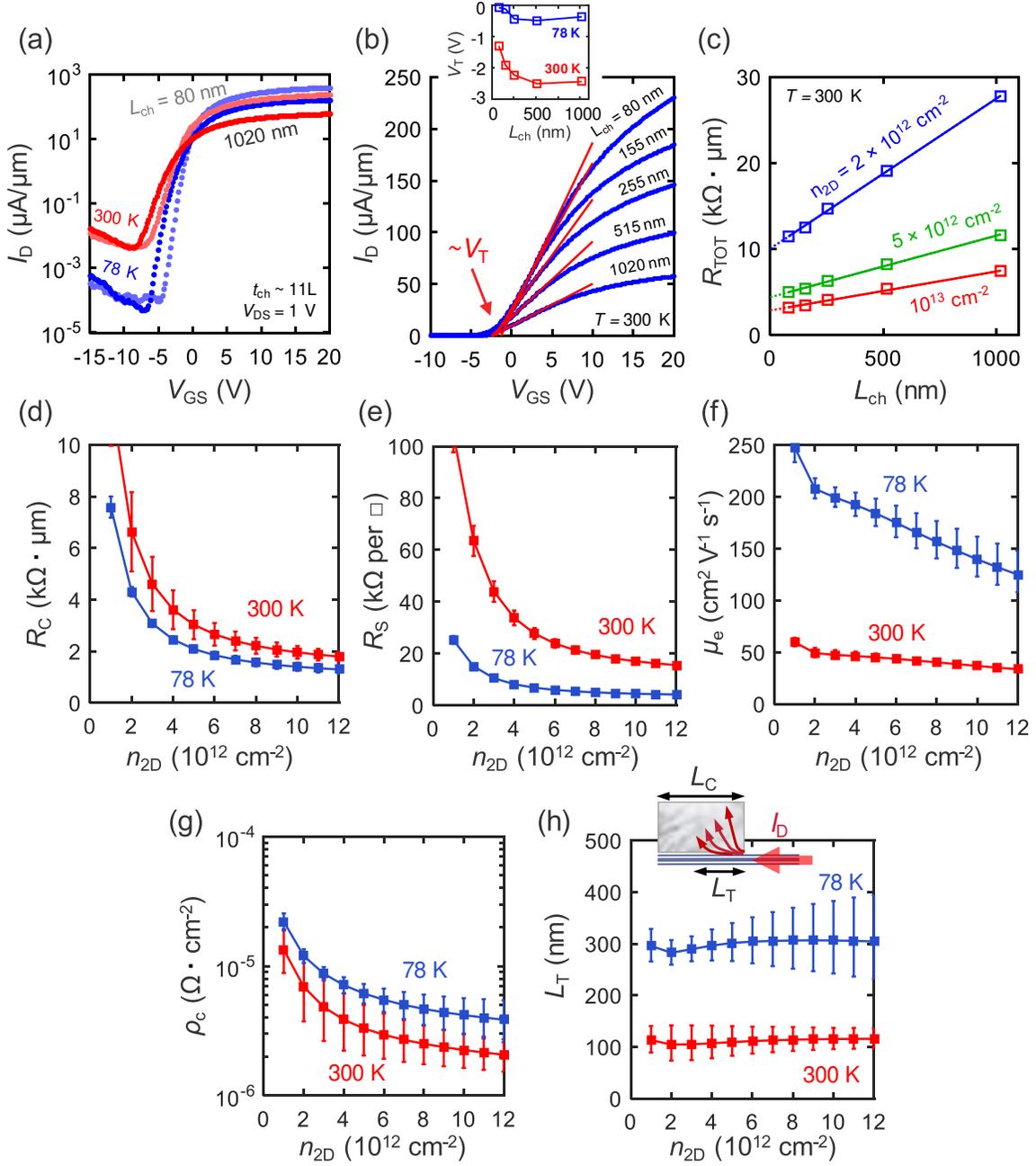

**Figure S6:** (a) Width-normalized $I_D$ vs. $V_{GS}$ characteristics of the longest and shortest channels of a 5-device, 11-layer MoTe$_2$ TLM test structure (Ag contacts, 30 nm SiO$_2$ back-gate). (b) Threshold $V_T$ extraction by extrapolating the linear $I_D$ regime to the $V_{GS}$-intercept. Inset shows $V_T$ versus channel length $L_{ch}$ at temperatures $T$ = 78 K and 300 K. (c) Width-normalized resistance $R_{TOT}$ vs. channel length with linear fits at varying channel carrier densities. $R_{TOT}$-axis intercept corresponds to $2R_C$ and the slope is sheet resistance $R_S$. Extracted (d) $R_C$ and (e) $R_S$ for this structure at 78 K and 300 K, demonstrating saturation of both parameters at high induced charge density (large gate overdrive). (f) Intrinsic electron mobility, calculated from $R_S$ and the 2D carrier density. (g) Specific contact resistivity $\rho_C$ and (h) transfer length $L_T$ vs. carrier density, at both temperatures. Inset graphic of (g) demonstrates the decay of current injection (red arrows) across $L_T$, ideally only a portion of the total contact length $L_C$.

## 5. Ag Contacts Deposited at Ultra-high Vacuum (UHV)

To fully elucidate the role of reactor pressure on Ag contacts to MoTe$_2$, we fabricated devices with pure 40 nm Ag contacts deposited at rates exceeding 2 Å/s in a custom-built UHV metal evaporation chamber, at pressures of 5×10$^{-9}$ Torr and below. **Figure S7** presents SEM micrographs (**Figure S7a**) and device characteristics (**Figure S7b-e**) of an encapsulated multilayer device following this process. A confluence of lower pressures, fewer incidental O- and H$_2$O- species, and a lack of sample rotation in a line-of-sight e-beam evaporator contributed to poor metal nucleation across the 30 nm SiO$_2$ substrate. This is evidenced in SEM profiles (**Figure S7a**) with localized discontinuities and agglomerations across the Ag leads. In contrast, all metal contacting MoTe$_2$ surfaces is continuous and more uniformly dispersed, indicating improved metal wettability and hinting at the interfacial chemical reactions outlined in following sections.

Despite metal constrictions potentially increasing lead resistance, TLM extractions produced room temperature contact resistances matching the range of high-field $R_C$ values for contacts evaporated at one to three orders of magnitude higher pressure (contrast **Figure S7d** with **Figure 3a** of the manuscript). No measurable improvement in $R_C$ was found at 80 K. Short-channel saturation current densities (**Figure S7c**) are consistent with those of non-UHV devices. Room temperature current densities in a $L_{ch}$ ~ 240 nm device reached ~230 µA/µm at $V_{DS}$ = 2.5 V, rising to ~350 µA/µm at $T$ = 80 K. Shorter channel length patterning was limited by shadowing from a lack of sample rotation during deposition.

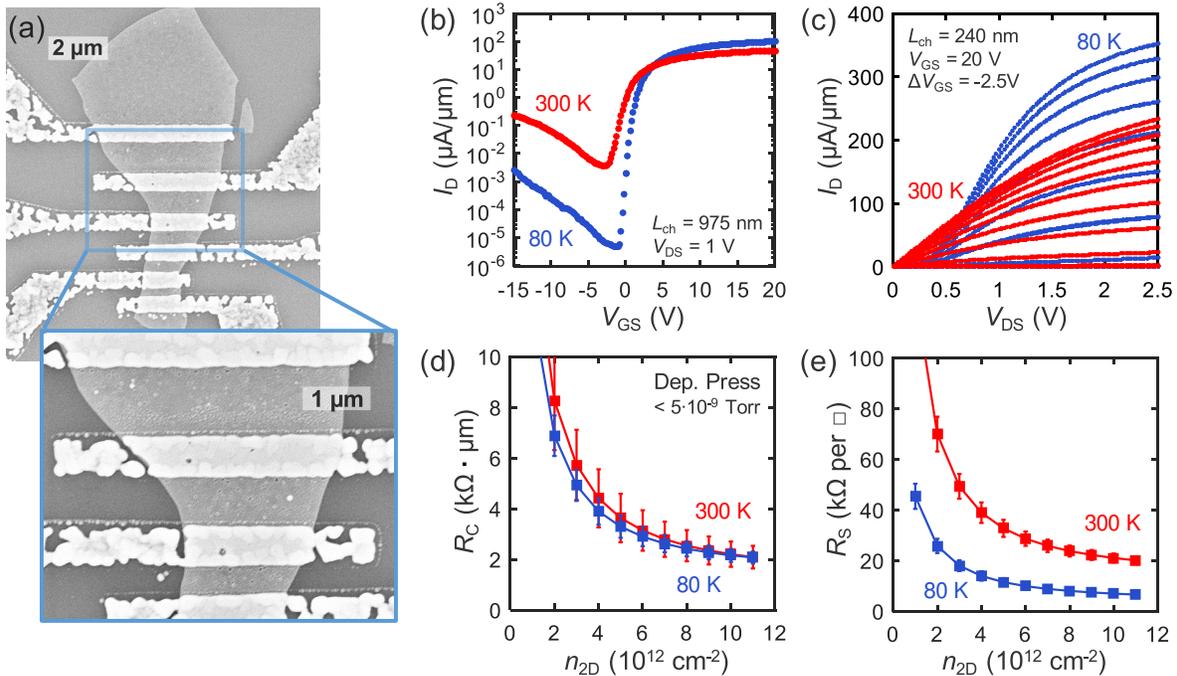

**Figure S7:** (a) SEM micrographs of an AlO$_x$ capped MoTe$_2$ FET on a 90 nm SiO$_2$ back gate, with nominally 40 nm-thick Ag contacts deposited under UHV ambient. Note the Ag metal leads poorly nucleate on SiO$_2$, whereas the Ag uniformly wets and deposits across the MoTe$_2$ surface. (b) $I_D$ vs. $V_{GS}$ transfer characteristics for the longest channel ($L_{ch}$ ~ 975 nm) at 80 K and 300 K ambient. (c) $I_D$ vs. $V_{DS}$ dual sweeps presenting short-channel saturation currents at 80 and 300 K. TLM-extracted (d) contact resistance and (e) sheet resistance for the pictured device, matching high-field values of devices fabricated under HV ambient (as shown in the manuscript).

## 6. Interface Chemistry Analysis as a Function of Deposition Chamber Ambient

All electronegativity values below are reported according to the Pauling electronegativity scale and all standard Gibbs free energy of formation ($\Delta G^{\circ}_f$) values are reported per chalcogen or oxygen atom or single -OH group.

### 6.1. Ag–MoTe₂

The chemical states found at 227.94 and 40.11 eV in the Mo 3*d* and Te 4*d* core level spectra after exfoliation (not shown) and also following Ag deposition at room temperature (RT) are indicative of metallic behavior of a small concentration of MoTe₂ within the probed region, associated with Te excess typically observed in MoTe₂ crystals.[21] These chemical states persist after Ag deposition regardless of reactor base pressure and are denoted in **Figures S8a** and **S8b** by '**'. No chemical states appear in either the Mo 3*d* or Te 4*d* core level spectra following Ag deposition under either ultra-high vacuum (UHV, base pressure

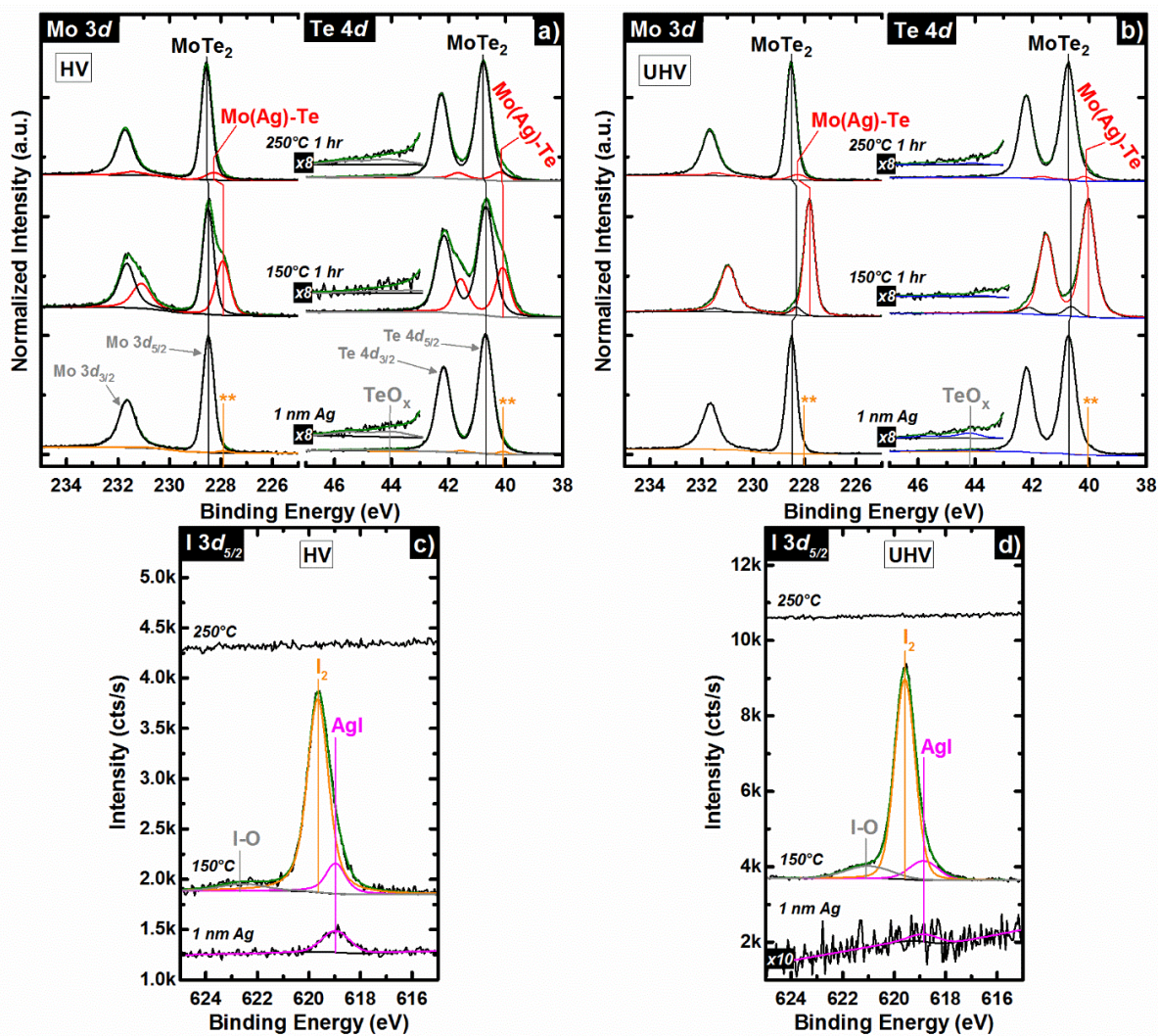

**Figure S8:** Mo 3*d*, Te 4*d*, and I 3*d₅/₂* core level spectra obtained from MoTe₂ after depositing 1 nm Ag under HV (a,c) and UHV (b,d) ambient and also following subsequent post metallization annealing at 150 °C and 250 °C.

< 2 × 10⁻⁹ mbar) or high vacuum (HV, base pressure < 5 × 10⁻⁶ mbar) conditions that would suggest the formation of reaction products. In addition, the bulk $MoTe_2$ chemical states in these core levels do not exhibit shifts which would suggest Ag-induced band bending, which is in contrast with strong Fermi level pinning near the charge neutrality level of $MoTe_2$ (4.77 eV) recently reported in the cases of other contact metals.[17] However, the devices discussed here are fabricated from multilayer $MoTe_2$ whereas C. Kim *et al.*[17] observe strong Fermi level pinning in back gated, single layer $MoTe_2$ devices. Therefore, direct comparison of contact performance may not be appropriate. In addition, a small concentration of native $TeO_x$ is detected on exfoliated $MoTe_2$ (not shown) and persists throughout Ag deposition and post metallization annealing as evidenced by the chemical state detected between 43.70 and 44.13 eV in the Te 4*d* spectra obtained throughout the experiment.

Following both Ag and Sc deposition, no evidence of e-beam related surface damage was observed across $MoTe_2$ spectra, consistent with prior studies of comparable contact deposition on $MoS_2$ and $WSe_2$.[22, 23]. The penetration depth of the most energetic characteristic X-rays of both metals exceeds 10 µm, far deeper than the <10 nm XPS probe depth and the thickness of relevant FET devices. Additionally, metal evaporation for device fabrication and XPS samples were all performed at relatively low rates at or below 0.1 nm/s, far lower than the rates (and resultant X-ray fluxes) associated with e-beam induced damage in thick gate oxides and underlying semiconductor channels.[24-26]

Annealing under UHV at 150 °C drives reactions between Ag and $MoTe_2$ resulting in the formation of a substantial concentration of intermetallic $Mo_xAg_{1-x}Te$ as evidenced by the high intensity chemical states at 227.86 and 40.06 eV regardless of reactor base pressure. $Mo_xAg_{1-x}Te$ interfaces with $MoTe_2$, while unreacted metallic Ag lies on the surface according to comparisons of relevant core level spectra obtained at two different takeoff angles (not shown here). The $Mo_xAg_{1-x}Te$ chemical states detected at low binding energy from the bulk $MoTe_2$ states in the Mo 3*d* and Te 4*d* core level spectra following the 250 °C anneal decrease in intensity by factors of 4 and 16 in the cases of Ag deposited under HV and UHV, respectively, presumably due to loss by desorption. This likely does not occur in the devices discussed here as they are fabricated with 40–60 nm thick Ag contacts and capped with a 20 nm $AlO_x$ film, which would both prevent contact metal desorption (as is observed in interface chemistry study with 1 nm Ag film employed) and limit perturbations at elevated temperatures to intermixing between Ag, I, and $MoTe_2$. Much of the deposited Ag and related reaction products ($Mo_xAg_{1-x}Te$, AgI) are desorbed in-situ during annealing. This is possibly due to the thermodynamically favorable formation of a substantial concentration of AgI during 150 °C anneal and desorption during 250 °C anneal made possible by the extremely thin Ag film.

Iodine, originating from the growth process, is present in a concentration near the limit of XPS detection (~0.1 atomic %) following Ag deposition at RT under UHV conditions (**Figure S8c**). Post metallization annealing at 150 °C results in the accumulation of a substantial amount of iodine within the vicinity of the Ag–MoTe₂ interface. The residual carrier gas primarily retains its original chemistry as evidenced by the binding energy of the most intense chemical state (619.67 eV),[27] however the elevated temperature drives increased concentration of AgI and the formation of bonds between iodine and oxygen based adsorbates (621.35 eV) presumably at the Ag–MoTe₂ interface.[28] Interestingly, iodine is below the limit of detection following annealing at 250 °C suggesting this step is enough to cause substantial iodine out-diffusion from the sample. The same is expected in the devices discussed here as they experience a similar thermal history to the samples fabricated for interface chemistry analysis. The evolution of chemical states associated with iodine detected in the I $3d_{5/2}$ core level spectra obtained from the second MoTe$_2$ sample throughout Ag deposition under HV and subsequent post metallization annealing (**Figure S8d**) are analogous with those shown in **Figure S8c**.

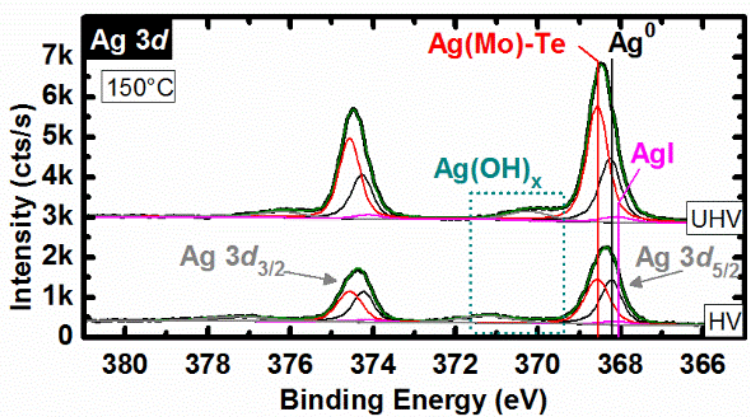

**Figure S9:** Ag $3d$ core level spectra from MoTe$_2$ metallized with 1 nm Ag under UHV and HV conditions (spectra labeled accordingly) and subsequently annealed at 150 °C under UHV and HV conditions (separate samples).

Annealing at 150°C catalyzes reactions between Ag and MoTe$_2$, drives substantial formation of Ag–I bonds, and facilitates I migration to the Ag–MoTe$_2$ interface. Iodine is presumably drawn to the Ag–MoTe$_2$ interface during annealing at 150°C as the intensities of the AgI chemical state in the 'UHV' and 'HV' Ag $3d$ core level spectra (**Figure S9**) exhibit marked increases compared with that detected following Ag deposition at room temperature. It is possible that the formation of intermetallic Ag$_{1-x}$Mo$_x$Te is catalyzed by I$_2$ migration to the Ag–MoTe$_2$ interface at elevated temperature, where chemical perturbation of the interface increases in severity with increased concentration of migrating I$_2$. The concentration of Ag$_{1-x}$Mo$_x$Te relative to metallic Ag under UHV conditions is 3.4:1 while that under HV conditions is 1.1:1. This seemingly coincides with 2.7× higher I$_2$ concentration detected in the 'UHV' sample compared with that in the 'HV' samples after annealing at 150 °C. Interfacial Ag–OH bonds formed upon Ag deposition are not affected by annealing as the intensity of corresponding chemical states in the Ag $3d$ core level spectra do not change upon annealing relative to the total Ag $3d$ intensity.

### 6.2. Sc–MoTe$_2$

MoTe$_2$ employed here exhibits slightly less severe Te excess, with Te:Mo ratio of 2.2 as indicated by the bulk MoTe$_2$ states detected from the exfoliated sample. Low binding energy states denoted by '**\*\***' in both the Mo $3d$ and Te $4d$ core level spectra (228.08 and 40.15 eV, respectively) obtained from exfoliated MoTe$_2$ correspond with metallic states associated with Te excess.[21]

Sc aggressively reacts with MoTe$_2$ when deposited under UHV conditions, completely reducing the top-most MoTe$_2$ to form metallic Mo and ScTe$_x$ as evidenced by the chemical states in addition to those of bulk MoTe₂, which appear in the Mo 3*d* and Te 4*d* core level spectra following deposition (227.79 and 40.39 eV, respectively; **Figure S10a**). Annealing at 150°C drives dissociation of Sc–Te bonds and subsequent formation of more energetically favorable Sc–O and Sc–OH bonds. The doublet at low binding energy in the Te 4*d* core level spectra obtained following annealing at 150°C and 250°C is a convolution of the Te$^{x+}$ state in both ScTe$_x$ and MoTe$_{x<2}$. MoTe$_{x<2}$ can form during annealing steps by Sc–Te bond disso-ciation and subsequent bond formation between Mo⁰ and Te$^{x+}$.

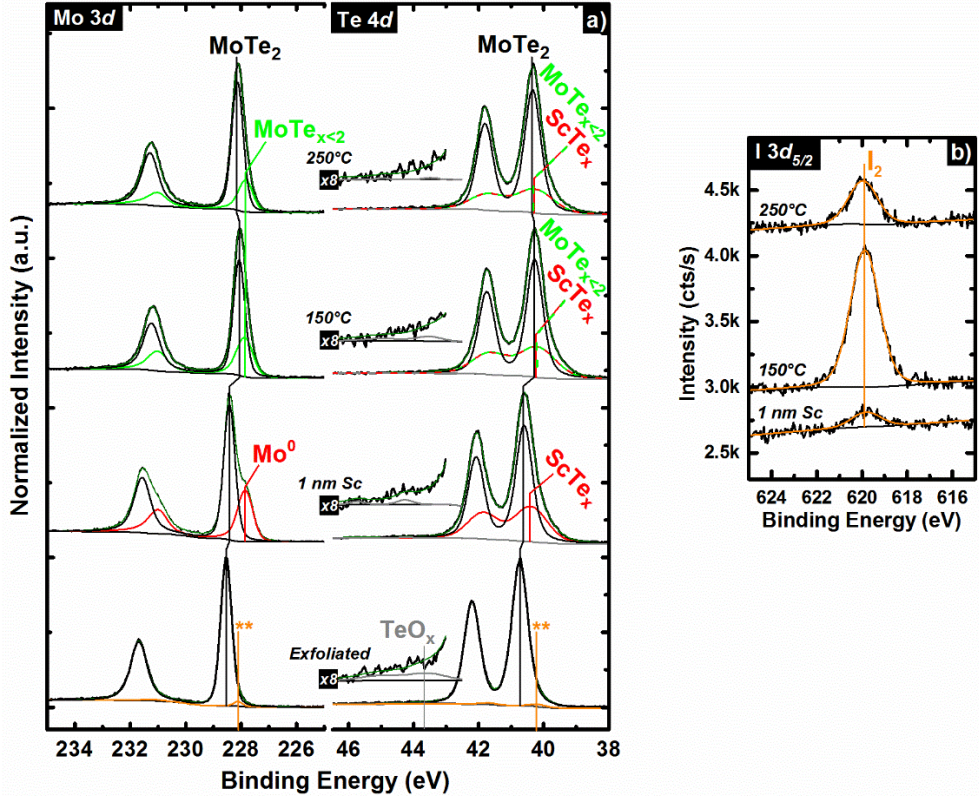

**Figure S10:** a) Mo 3*d* and Te 4*d* core level spectra obtained from exfoliated MoTe₂. a) Mo 3*d*, Te 4*d*, and b) I 3$d_{5/2}$ core level spectra obtained from MoTe₂ after 1 nm Sc deposition at ~300 K under UHV conditions and following post metallization annealing at 150 °C and 250 °C.

Interestingly, interactions between Sc and iodine are below the limit of detection by XPS following either deposition at RT or post metallization annealing (**Figure S10b**). Residual iodine is detected at 619.80 eV following Sc deposition. Similar to Ag contacts, annealing at 150 °C drives iodine aggregation within the vicinity of the Sc–MoTe₂ interface. However, dissimilar to Ag contacts, no additional chemical states in the I 3$d_{5/2}$ core level spectrum are detected aside from that representing I₂. This suggests reactions between Sc and excess Te in MoTe₂ or adsorbed gaseous species on the MoTe₂ surface are more energetically favorable than the formation of Sc–I bonds. Sc getters the adsorbed species initially present on the MoTe₂ surface, preventing the formation of I-O bonds, which are observed at the Ag–MoTe₂ interface following annealing at 150 °C. Interestingly, the concentration of iodine remains above the limit of XPS detection following annealing at 250 °C unlike in the Ag–MoTe$_2$ case. It is possible that the diffusion rate of iodine through the Sc$_x$O$_y$/Sc(OH)$_x$ film formed *in–situ* is far slower than that through Ag.

### 6.3. XPS Experimental Details

*Interface chemistry at room temperature:*

Bulk MoTe$_2$ flakes were mechanically exfoliated via Scotch Tape method. Ag and Sc source material with 99.99% purity employed in this work was purchased from Kurt J. Lesker.

*Interface chemistry formed under ultra–high vacuum (UHV, 5 × 10$^{-9}$ mbar) conditions:*

Exfoliated MoTe$_2$ was loaded into a UHV cluster tool (base pressure 10$^{-9}$ mbar, described elsewhere)[29] as quickly as possible (<5 min air exposure) and the initial surface was characterized by X-ray photoelectron spectroscopy (XPS). Metal source outgassing and deposition to a thickness of 1 nm was performed using the same procedure as employed in similar studies and described elsewhere.[23] The temperature of the sample did not exceed 30°C throughout the deposition process. The configuration of the deposition chamber requires the deposition rate to be determined with quartz crystal monitor prior to deposition on MoTe$_2$.

*Interface chemistry formed under high vacuum (HV) conditions:*

The sample was exfoliated and immediately loaded into the cleanroom deposition tool (Temescal BJD–1800 e-beam evaporator). The metal source outgassing and deposition to a thickness of 1 nm (base pressure 5 × 10$^{-6}$ mbar) was performed using the same procedure as has been employed in similar studies and described elsewhere.[23] The thickness of the Ag film was monitored *in situ* by quartz crystal monitor. After shutting off the e-beam, the sample was transferred as quickly as possible (<5 min air exposure) from the cleanroom deposition chamber to UHV cluster tool for XPS.

*Post metallization annealing*

After characterizing the interface chemistry of the samples fabricated at RT with XPS, each respective sample was transferred to the deposition chamber attached to the UHV cluster tool without breaking vacuum and annealed at 150°C and 250°C under UHV conditions for 1 hour each. Following each annealing step, the interface chemistry was characterized *in-situ* by XPS.

*X-ray Photoelectron Spectroscopy*

A monochromated Al Kα source (1486.7 eV), takeoff angle of 45° and Omicron EA125 hemispherical analyzer with ±0.05 eV spectral resolution, pass energy of 15 eV, and acceptance angle of 8° were employed during spectral acquisition. The analyzer was calibrated according to ASTM E2108.[30] AAnalyzer was employed to deconvolve all spectra.[31]

## 7. Saturation Current Density

As discussed in the manuscript, cooling to 77–80 K yields reduced contact resistance and enhanced mobility, subsequently increasing saturation current density in Ag-contacted *n*-type transistors by 50-100%. **Figure S11** presents sample $I_D$ vs. $V_{DS}$ sweeps for two short channel devices in this temperature range, representing both a thin-channel sample (**Figure S11a**; 5-layer, $t_{ch} \approx 3.5$ nm) and one approaching the bulk limit (**Figure S11b**; 17-layer). Saturation current densities at ~80 K ambient increase from approximately 350 to 450 μA/μm across this thickness range. In the thicker sample in **Figure S11b**, the sub-linear $I_D$ vs. $V_{DS}$ at $V_{DS} < 0.5$ V is likely due to increased access resistance from interlayer screening.

## 8. h-BN Growth and Device Integration

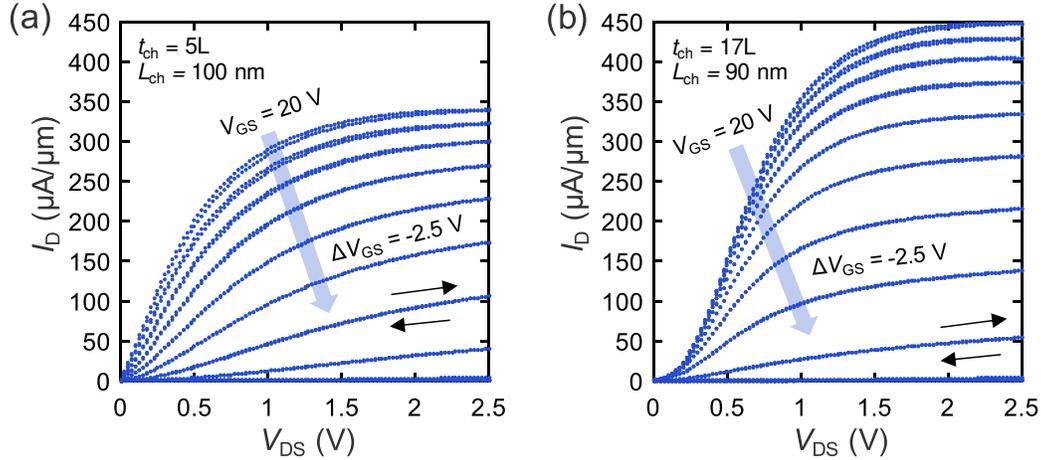

**Figure S11:** $I_D$ vs. $V_{DS}$ dual sweeps of (a) 5- and (b) 17-layer, Ag-contacted MoTe$_2$ short-channel FETs on a 30 nm SiO$_2$ back-gate at 78–80 K, demonstrating record saturation current densities increasing as channel thickness approaches bulk limits.

We grew continuous, large-area hexagonal boron nitride (h-BN) monolayers on re-usable Pt foils in a 2 inch furnace-based low-pressure chemical vapor deposition (CVD) process, using a borazine / hydrogen flux derived through thermal decomposition of an ammonia borane (BH$_3$NH$_3$) solid precursor.[32] During the growth, the furnace temperature is set to 1100°C while the precursor is heated to 100°C, facilitating the decomposition of ammonia borane into hydrogen, polyaminoborane, and borazine. Hydrogen gas is flowed through the furnace so that the borazine gas can diffuse to the platinum foil and adsorb onto the surface.

An electrochemical bubbling method is used to transfer the h-BN to the target substrate, so that the polycrystalline Pt growth substrate (15 mm x 24 mm) can be reused. A PMMA layer is spin coated onto the h-BN while still on the Pt foil, followed by a layer of polystyrene (PS). The PMMA is primarily for adhesion to the h-BN, while the PS is a more rigid polymer layer for ease of handling. The entire structure (PS/PMMA/h-BN/Pt) is attached to an electrode and submerged into a 1 M NaOH solution with a Pt mesh also submerged as the anode. Applying a voltage difference between the Pt foil substrate and Pt mesh induces H$_2$ bubbles, which delaminate the PS/PMMA/h-BN from the Pt.

Centimeter-scale h-BN films were subsequently transferred onto exfoliated MoTe$_2$ using these transfer stamps, applied directly on an 80°C hot-plate, necessitating several minutes of flake exposure to ambient atmosphere. Transfer-stamp adhesion was achieved through a series of 80-130°C bakes, on both hot-plates and in low-vacuum ovens, prior to polymer removal through a 2-hour, glovebox-based soak in N-methyl-2-pyrrolidone (NMP; commercially as Remover PG resist stripper). A final solvent dip was followed by a 1–3 hour-long, 200°C anneal in $N_2$ ambient, prior to device fabrication with Sc/Cr/Ag top contacts as described in the manuscript. Select samples were subject to an hour-long, 400°C high-vacuum anneal, which improved the yield of clean, measurable channels (though not device performance).

**Figure S12** presents optical images, AFM micrographs and Raman spectra of transferred monolayer h-BN onto $SiO_2$/Si substrates. The presence of monolayer h-BN on 285 nm $SiO_2$ is confirmed by the Raman peak at 1369 cm$^{-1}$. The characteristic h-BN peak is centered at approximately 1366 cm$^{-1}$ in bulk h-BN, but will exhibit blue shifts up to 4 cm$^{-1}$ in monolayer, due to a hardening of the $E_{2g}$ phonon mode.

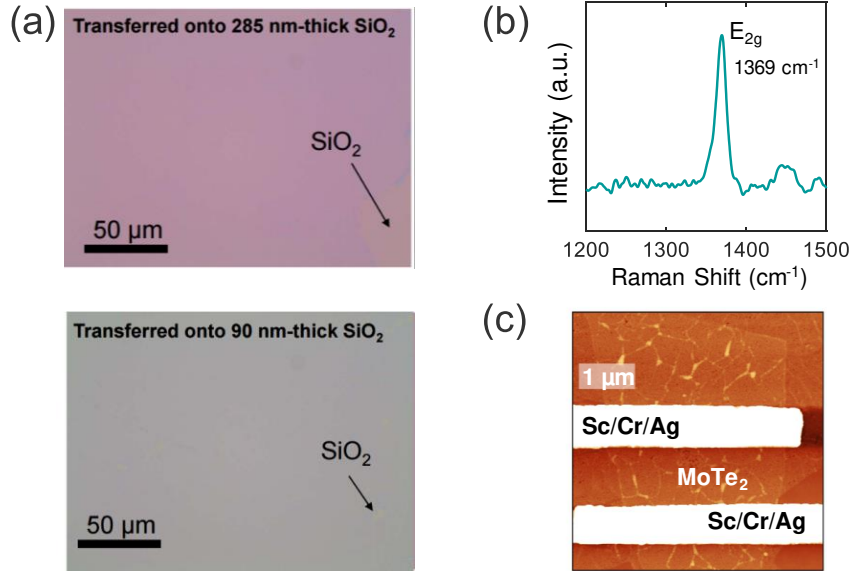

**Figure S12:** (a) Optical images of dry-transferred monolayer h-BN on 285 nm (top) and 90 nm (bottom) $SiO_2$/Si substrates. Arrows indicate large tears and gaps in an otherwise uniform mm-scale film. (b) Raman spectra of transferred monolayer h-BN on 285 nm $SiO_2$ (532 nm laser). (c) AFM-micrograph of an $AlO_x$/h-BN capped $MoTe_2$ device depicting complete coverage and some wrinkling/transfer residue across the encapsulating monolayer.

**Figure S13** displays a representative TEM cross-section of a thin few-layer $MoTe_2$ device with h-BN/Sc MIS contacts. We observe consistent layer count between $MoTe_2$ channel and contact regions, with the h-BN monolayer distinct in both images. This transferred layer maintains a sizeable van der Waals gap above the top-most $MoTe_2$ layer within the device channel. Underneath the contacts, however, this gap is apparently reduced between the top 2–3 layers due to the pressure from the deposited metal stack (with each of the Sc/Cr/Ag layers evaporated at rates matching or exceeding 2 Å/s in the UHV reactor). We note no metal-telluride intermixing with only 7 layers of semiconductor; this can be compared to the electrical data from **Figure 4a** of the manuscript, where lack of conduction across 5-layer flakes with *direct* Sc contacts suggests local consumption of all $MoTe_2$ at those contact regions.

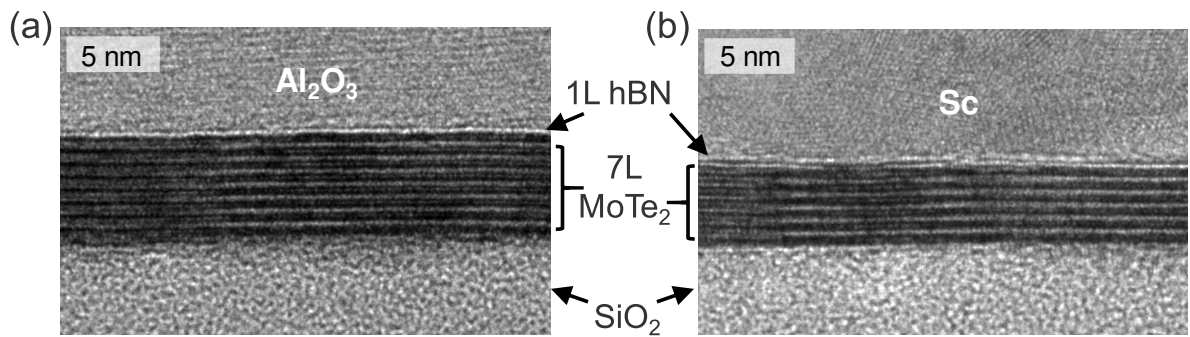

**Figure S13:** TEM cross-section of the (a) channel and (b) MIS contact regions of a 7-layer $MoTe_2$ transistor capped with monolayer h-BN. Consistent layer counts between both regions confirm the role of this monolayer as a diffusion barrier, preventing intermixing between the layered semiconductor and the Sc contact metal.

## Supplementary References